\documentclass[prd,twocolumn,reprint,aps,superscriptaddress,nofootinbib]{revtex4-1}

\usepackage{graphicx}
\graphicspath{{Figures/}}
\usepackage{dcolumn}
\usepackage{bm}        
\usepackage{amssymb}   
\usepackage{mathtools}  
\usepackage[svgnames]{xcolor}
\usepackage[caption=false]{subfig}
\usepackage{multirow}
\usepackage[inline]{enumitem}
\usepackage{microtype}

\usepackage{hyperref}
\usepackage[separate-uncertainty=true]{siunitx}
\hypersetup{
        colorlinks=true,
        citecolor=MediumPurple,
        filecolor=LimeGreen,
        linkcolor=RoyalBlue,
        urlcolor=MediumPurple
}
\usepackage{cleveref}
\Crefname{equation}{Eq.}{Eqs.} 

\begin{document}

\title{Scattering Loss in Precision Metrology due to Mirror Roughness}

\author{Yehonathan Drori}
\affiliation{LIGO Laboratory, California Institute of Technology, Pasadena, CA 91125, USA}

\author{Johannes Eichholz}
\affiliation{LIGO Laboratory, California Institute of Technology, Pasadena, CA 91125, USA}
\affiliation{Centre for Gravitational Astrophysics, College of Science, The Australian National University, Canberra, ACT 2600, Australia}

\author{Tega Edo}
\affiliation{LIGO Laboratory, California Institute of Technology, Pasadena, CA 91125, USA}
\affiliation{Department of Physics and Astronomy, The University of Sheffield, S3 7RH, UK}

\author{Hiro Yamamoto}
\affiliation{LIGO Laboratory, California Institute of Technology, Pasadena, CA 91125, USA}

\author{Yutaro Enomoto}
\affiliation{LIGO Laboratory, California Institute of Technology, Pasadena, CA 91125, USA}
\affiliation{Department of Applied Physics, School of Engineering, The University of Tokyo, 7-3-1 Hongo, Bunkyo-ku, Tokyo 113-8656, Japan}

\author{Gautam Venugopalan}
\affiliation{LIGO Laboratory, California Institute of Technology, Pasadena, CA 91125, USA}

\author{Koji Arai}
\author{Rana X Adhikari}

\begin{abstract}
Optical losses degrade the sensitivity of laser interferometric instruments. They reduce the number of signal photons and introduce technical noise associated with diffuse light. In quantum-enhanced metrology, they break the entanglement between correlated photons. Such decoherence is one of the primary obstacles in achieving high levels of quantum noise reduction in precision metrology.

In this work, we compare direct measurements of cavity and mirror losses in the Caltech 40m gravitational-wave detector prototype interferometer with numerical estimates obtained from semi-analytic intra-cavity wavefront simulations using mirror surface profile maps. We show a unified approach to estimating the total loss in optical cavities (such as the LIGO gravitational detectors) that will lead towards the engineering of systems with minimum decoherence for quantum-enhanced precision metrology.

\end{abstract}

\maketitle

\twocolumngrid

\section{Introduction}
\label{s:intro}
Laser interferometry is a pillar of modern precision measurement and plays an integral role in the preparation, interrogation, and manipulation of quantum systems, including entangled states of light and matter. Optical losses pose both practical and fundamental limitations to instrument performances. Absorption and scatter from optical components or subsystems decrease classical signal levels and introduce application-dependent technical noise. Optical losses also drive the decoherence of non-classical photon states, which find application in quantum computing, cryptography, and quantum-enhanced metrology. A typical example for the latter is the use of \emph{squeezed light} in ground-based interferometric gravitational-wave (GW) detectors to reduce photon shot noise~\cite{Caves1981}. We discuss optical loss as a source of decoherence for quantum states in the context of these quantum noise-limited instruments.

The ability to directly observe gravitational waves (GWs) is continuously evolving our understanding of the universe and its history. The number of observed mergers of binary systems of black holes and neutron stars is steadily growing~\cite{gwtc1,gwtc2}, and keeps informing black hole population models, formation channels for compact objects, and parameter constraints on theories of gravity and the equation of state of highly degenerate nuclear environments. Improved sensitivity, for example from a reduction of quantum noise, and detections of other, new source types promise to multiply this knowledge across the GW frequency spectrum. 

Interferometric GW detectors are electromagnetic transducers for GWs that convert the strain of passing GWs on space-time to phase modulations of a laser carrier. The two Advanced LIGO (Adv. LIGO) observatories~\cite{aasi2015_advanced-ligo}, Advanced Virgo (AdVirgo)~\cite{acernese2015_advanced-virgo}, KAGRA~\cite{aso2013_kagra}, and GEO600~\cite{affeldt2014_geo600} are ground-based detectors that are sensitive to signals in the audible frequency range. They all have similar optical topologies, using a combination of optical cavities to increase the number and storage time of photons interacting with passing GWs.
Optical loss generates a variety of technical problems for the control of these detectors. It broadens cavity line widths, affects the behavior of coupled-cavity systems, and back-scattering can create transient signal glitches or pollute entire frequency bands~\cite{Vinet_1996_ScatterCoherentEffects,Vinet_1997_ScatterStatistics,Ottaway_2012_Upconversion}. Radiation pressure-induced opto-mechanical effects due to scattered fields may also become relevant. A thorough understanding of the magnitude and nature of optical loss will help anticipate and mitigate such effects.

During their third observation run (O3), the measured round-trip losses of the Fabry-Perot arm cavities of the Adv. LIGO detectors 
were in the range 60\,--\,70\,ppm. This is below the requirement of 75\,ppm and not currently considered a limiting factor for the interferometer performance. The optical absorption of the high-quality dielectric mirror coatings is below the ppm level, which makes scatter the dominant loss mechanism in the interferometers. The expected arm loss based on surface figure phase maps and large angle scattering data is about 20\,--\,30\,ppm less than the measured value. Part of this discrepancy stems from the scattering into the angular region $\sim0.1$\,--\,$1^{\circ}$, which is difficult to access experimentally without ambiguity and not covered by existing measurements. It is therefore desirable and important to have a set of techniques that can reliably, repeatably and accurately characterize optical loss, so that mitigation efforts can be validated.

This work examines loss measurements at the Caltech 40\,m interferometer (40m), which is a small-scale prototype with the same principal optical topology as Adv. LIGO, and compares them with simulations to verify these models for the purpose of extending the analysis to other optical systems. The paper is structured as follows: \Cref{s:theory} provides necessary formulas and quantitative and qualitative description of the scattering and loss. \Cref{s:Measurements} describes the optical measurements utilized to estimate the total optical cavity losses. Finally, \Cref{s:Comparison} compares the simulations with the measurements and discusses the remaining discrepancies.

\section{Theory and Model of Scatter Loss}
\label{s:theory}
\begin{figure}[h]
    \centering
    \includegraphics[width=\columnwidth]{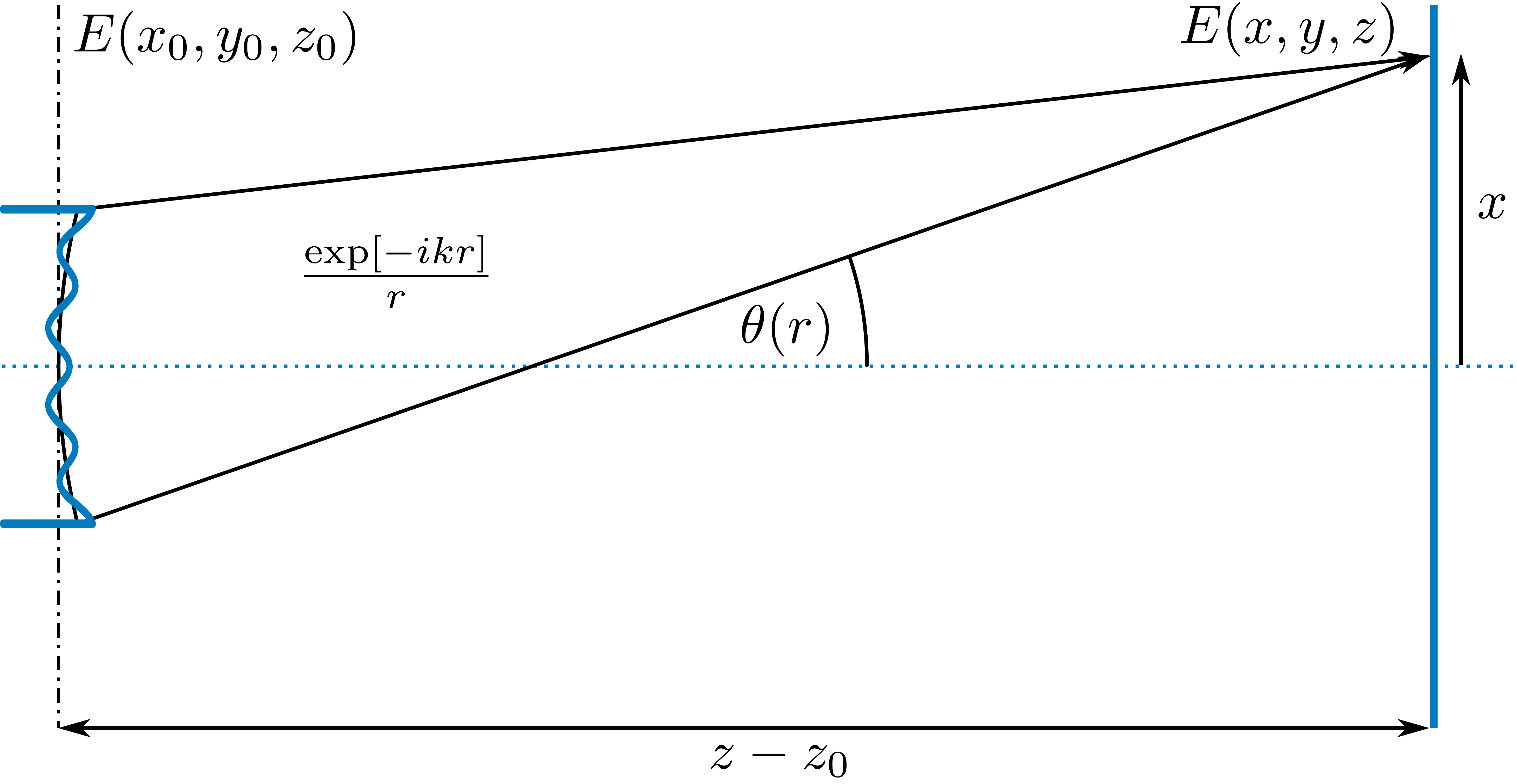}
    \caption{An illustration of scattering using the Fresnel integral in \Cref{eq:Huygen}. The source field, $E(x_0,y_0,z_0)$, derives from the reflections off the surface of a curved mirror---with surface aberration, $h$, defined as the deviation of the real mirror surface from a spherical surface. The sink field, $E(x,y,z)$ is the weighted sum of the source field, with a weighting factor that comprises two terms, a spherical wave, $\exp[-ikr]/r$, that encodes amplitude and phase interference information from the different locations on the mirror surface, and the obliquity factor, $\cos(\theta)$, that encodes the departure of the scattering vector from the optical axis, in accordance with Huygens principle.
    } 
    \label{fig:Huygen}
\end{figure}

In a long baseline Fabry-Pérot (FP) cavity, we wish to know the optical field distribution in the plane of the mirror, so that we can estimate the amount of power in the cavity using the field reflected from the mirror whilst also getting a good estimate of the power loss due to non-zero field values outside the mirror. To this end, we employ the Huygens-Fresnel integral
\begin{equation}
    \label{eq:Huygen}
    E(x,y,z)=\frac{i}{\lambda}\iint dx_0 dy_0 E(x_0,y_0,z_0) \frac{\exp[-ikr]}{r} \cos(\theta),
\end{equation}
where $E(x_0,y_0,z_0)$ and $E(x,y,z)$ are the source and sink field distributions, respectively, which will become the locations of the two mirrors, and we use the definitions $k=2\pi/\lambda$, $r^2 = \rho^2 + L^2$, $\rho^2 = (x-x_0)^2 + (y-y_0)^2$, $L=z-z_0$ and $\cos(\theta)=L/r$. 
A pictorial representation of this geometry is shown in \Cref{fig:Huygen}.

An eigenmode of \Cref{eq:Huygen} that is of particular interest in long-baseline FP cavities is the TEM$_{00}$ Gaussian beam mode, given by
 \begin{equation}
   E_q(x,y,z) = \sqrt{\frac{2 P_0}{\pi w^2_0}}\frac{q_{0}}{q(z)}\exp\bigg[-ik\bigg(z+\frac{x^2 + y^2}{2 q(z)}\bigg)\bigg],
   \label{eq:tem00}
\end{equation}
where $P_0$ is the total power in the beam, $w_0$ is the beam waist and $q(z)$ is the complex beam parameter---given by the expression
\begin{equation}
    q(z) = z - z_w + iz_R,
\label{eq:qParam}
\end{equation}
where $z_w$ is the location of the beam waist, $z_R$ is the Rayleigh range of the beam and $q_{0}=iz_R$; Or equivalently its reciprocal 
\begin{equation}
    \frac{1}{q(z)} = \frac{1}{R(z)} - i\frac{\lambda}{\pi w^2(z)}.
\label{eq:qParamInv}
\end{equation}
which is parameterized by the radius of curvature (RoC), $R(z)$, and the radius of the beam, $w(z)$.

\subsection{Surface Aberration and Scattering}
\label{sec:FieldScat}
The surface profile $h(x_0,y_0)$ of the mirror can be calculated from its Fourier transform $H(f_x,f_y)$:
\begin{equation}
h(x_0,y_0) = \iint H(f_x, f_y) \exp\big[-i 2 \pi (x_0 f_x + y_0 f_y)\big] \,df_x\,df_y
\label{eq:ASD}
\end{equation}
and the two dimensional power spectral density (PSD) is given by the expression
\begin{equation}
    PSD_2(f_x, f_y) \equiv |H(f_x, f_y)|^2.
\label{eq:PSD}
\end{equation}

When a Gaussian field is reflected by a mirror surface with height aberration $h$ and nominal radius of curvature $R_0$, the disturbance of the reflected field\,\footnote{The prompt reflected field at the mirror surface is $E(x_0,y_0,z_0)= \exp\big[ 2 i k \big(h + ({x_0^2 + y_0^2})/{2R_0}\big)\big] E_q(x_0,y_0,z_0)$.}, in the limit of small $h$ where $2kh\ll1$, is
\begin{equation}\label{eq:dE_src}
    \delta E(x_0,y_0,z_0) \simeq 2 i k h \exp\Big[ i k \frac{x_0^2 + y_0^2}{R_0}\Big] E_q(x_0,y_0,z_0).
\end{equation}
Note that in a resonating cavity, the radius of curvature of the beam impinging the mirror surface is $R(z_0) = R_0$.

We use \Cref{eq:Huygen} to propagate \Cref{eq:dE_src} to the second mirror. The wavefront mismatch at $z$ will determine the amount of power scattered out of the cavity mode by the first mirror.
The scattered field distribution in the $(x,y)$ plane displaced a distance $L$ from the mirror is 
\begin{equation}\label{eq:dE_sink}
    \delta E(x,y,z) = 2 i k \cdot M(\beta^{-1}x,\, \beta^{-1}y),
\end{equation}
where $\beta = \lambda (z-z_0)$. Here, the Fresnel integral has been recast as a convolution integral in frequency space, namely:
\begin{equation}\label{eq:dE_fft}
    M(\nu_x,\nu_y) = \eta \iint H(f_x, f_y) G_q(\nu_x-f_x,\nu_y-f_y)\,df_x\,df_y,
\end{equation}
where
\begin{equation}\label{eq:eta_fft}
    \eta(\nu_x,\nu_y) = \exp\big[{-i\pi\beta(\nu_x^2 + \nu_y^2)}/{2}\big] 
\end{equation} 
and
\begin{equation}\label{eq:G_fft}
    G_q(\nu_x,\nu_y) = \frac{E_q(\beta \nu_x, \, \beta \nu_y, \, z_0+L)}{\eta(\nu_x, \, \nu_y)}.
\end{equation}  

\begin{figure}[htbp]
\centering
    \includegraphics[width=\columnwidth]{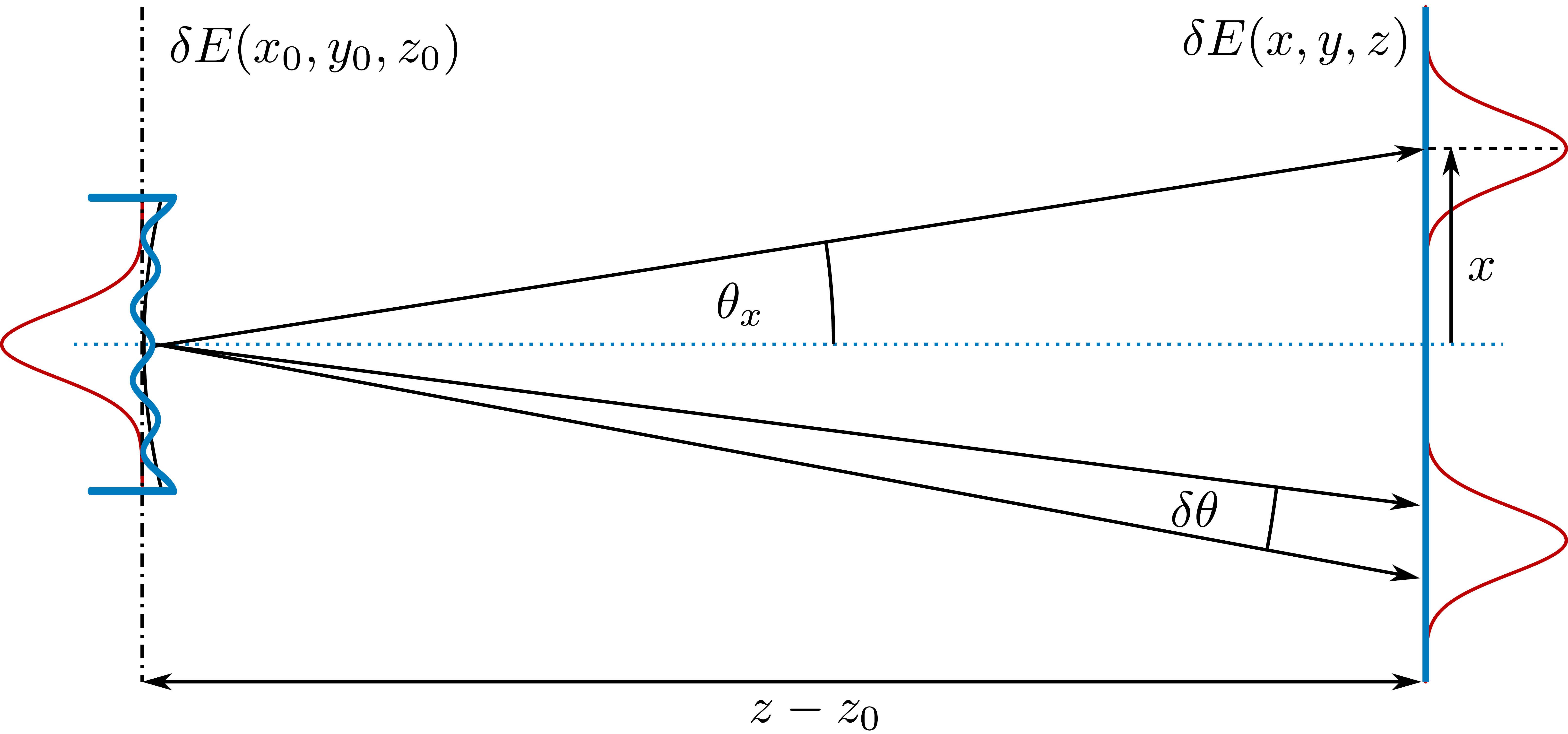}    
\caption{Scattering by continuous aberration. In the frequency domain, we have $H(f_x,f_y) \simeq 2 i k h_{\max}(f_s) \delta(f_x^2-f_s^2)$ for a continuous aberration of spatial frequency, $f_s$ along the $x-$direction. Thus, the field distribution in the sink plane is the sum of two displaced versions of $E_q$ with an amplitude scaling of $2 i k h_{\max}$.}
\label{fig:ASDscat}
 \end{figure}

The far field condition, satisfied by long baseline IFO, means that spatial coordinate and spatial frequency are related as follows:
\begin{equation}
    x = \beta \nu_x, \quad y = \beta \nu_y.
\label{eq:coord_freq}
\end{equation}
Furthermore, the angular coordinate, $\theta_{x,y}$, is related to spatial frequency, $\nu_{x,y}$, and spatial wavelength, $\lambda_{x,y}$, of the exitwave leaving the mirror surface via 
\begin{equation}
    \sin(\theta_{x,y}) = \lambda \nu_{x,y}, \qquad \sin(\theta_{x,y}) = \lambda/ \lambda_{x,y}
\label{eq:angle_freq}
\end{equation}
respectively. For small angles, this approximates to
\begin{equation}
    \theta_{x,y} \simeq \lambda \nu_{x,y}, \qquad \theta_{x,y} \simeq \lambda/ \lambda_{x,y}.
\label{eq:angle_freq_approx}
\end{equation}
\Crefrange{eq:dE_sink}{eq:angle_freq} show that an aberration with spatial wavelength $\lambda_{x,y}$ reflects the field towards the nominal angular direction $\theta_{x,y}$ with broadening, $\delta\theta = 2w(z)/L$, as illustrated in \Cref{fig:ASDscat}. 

The finite size of the mirrors defines a characteristic cutoff angle $\theta_{cut}$ and a corresponding spatial wavelength $\lambda_s^{cut}$ of surface deformations. All power scattered to angles larger than $\theta_{cut}$ by surface deformations with length scales smaller than $\lambda_s^{cut}$ is lost entirely, similar to the tail of the reflected fundamental Gaussian mode which misses the mirror. At angles smaller than $\theta_{cut}$, the overlap between impinging wavefront distortions and mirror surface needs to be evaluated to determine the power lost. 

For FP Michelson interferometers such as aLIGO and and the Caltech 40\,m prototype schematically shown in \Cref{fig:40m-config} we can define for each arm cavity an input mirror (IM), where the laser beam enters the cavity, and an end mirror (EM). Important geometrical parameters for the FP arm cavities of aLIGO and the Caltech 40\,m prototype that are used in this calculation are listed in \Cref{tab:FPparams} along with the parameters derived above. As can be seen, the large separation of the LIGO mirrors results in over a factor of 20 longer cutoff wavelength over the 40\,m's.
 
\begin{center}
	\begin{table}
		\renewcommand{\arraystretch}{1.5}\resizebox{\columnwidth}{!}
		{
        	\begin{tabular}{ @{\hspace{0.5cm}} c @{\hspace{0.5cm}}| @{\hspace{0.5cm}}c @{\hspace{0.5cm}}|@{\hspace{0.5cm}}c @{\hspace{0.5cm}} } \hline \hline
             Parameters & aLIGO  & 40\,m \\ \hline
             Cavity length $(L)$ & 3994.5\,m & 38\,m \\
             Mirror Aperture $(a)$ & 34\,cm & 7.5\,cm \\
             RoC of IM $(R_1)$ &  1934\,m &  flat \\ 
             RoC of EM $(R_2)$ & 2245\,m &  57.5\,m \\ 
             Beam radius on IM $(w_1)$ & 5.3\,cm & 0.30\,cm \\ 
             Beam radius on EM $(w_2)$ &  6.2\,cm & 0.52\,cm \\ 
             $\theta_{cut}\, (= a/2L)$  & $0.0024^{\circ}$ & $0.057^{\circ}$ \\ 
             $\lambda_s^{cut}\, (= \lambda / \theta_{cut})$  & 2.5\,cm & 1.1\,mm \\ 
             $\delta P_m/P \,\big(=\sum_j \exp\big[{-2 a^2/w_j^2}\big] \,\big)$ & 0.3\,ppm & $10^{-39}$\,ppm \\ \hline \hline
        	 \end{tabular}
         }
         \caption{Fabry-Perot cavity parameters for the aLIGO arms and the 40\,m arms. Here $a$ is the diameter of the mirror and $w_j$ is the half-width of the beam at the $j^{th}$ mirror.
		} 
         \label{tab:FPparams}
     \end{table}
 \end{center}

\subsection{BRDF as a loss estimation metric}
\label{sec:BRDF}
The bidirectional reflectance distribution function (BRDF) describes the angular dependence of the intensity of light reflected and scattered from the mirror surface \cite{Stover_2012}, defined by the following formula 
\begin{equation}
    \label{eq:BRDF}
	BRDF\left(\theta,\phi\right) =\frac{L^2}{P_0\cos(\theta)} |E(x,y,z)|^2,
\end{equation}
where $\theta$ and $\phi$ are angles in spherical coordinates, with the usual relation to Cartesian coordinate i.e. $x=L\sin(\theta)\cos(\phi)$ and $y = L\sin(\theta)\sin(\phi)$. Again, $P_0$ is the total power in the beam whilst $L$ is the distance between the mirror and the point of measurement.

The one-dimensional distribution used often for loss estimation is
\begin{equation}
\label{eq:BRDF_1D}
    BRDF(\theta) = \frac{1}{2\pi} \int_{0}^{2\pi} d\phi ~ BRDF(\theta, \phi),
\end{equation}
which is fairly accurate when no explicit dependence on $\phi$ is present in the phase map.

Direct BRDF measurements of a mirror are usually limited to above several degrees from the normal, because the small spatial separation from the incident beam hinders the measurement of the small scattered power. Therefore, the BRDF for small angles is best estimated using scattering calculations from the phase map $h(x_0,y_0)$ of the mirror surface by numerical methods. The angular range of the resulting BRDF data is limited by the spatial resolution of the phase map data, which is typically of order $\mathcal{O}$(0.1\,mm), so that the maximum scattering angle covered by phase map data, using \Cref{eq:angle_freq_approx}, is $\sim0.3^{\circ}$. This leaves a gap in the angular range between phase map estimations and direct BRDF measurements.

\begin{figure}[tbp]
\centering
		\includegraphics[width=\columnwidth]{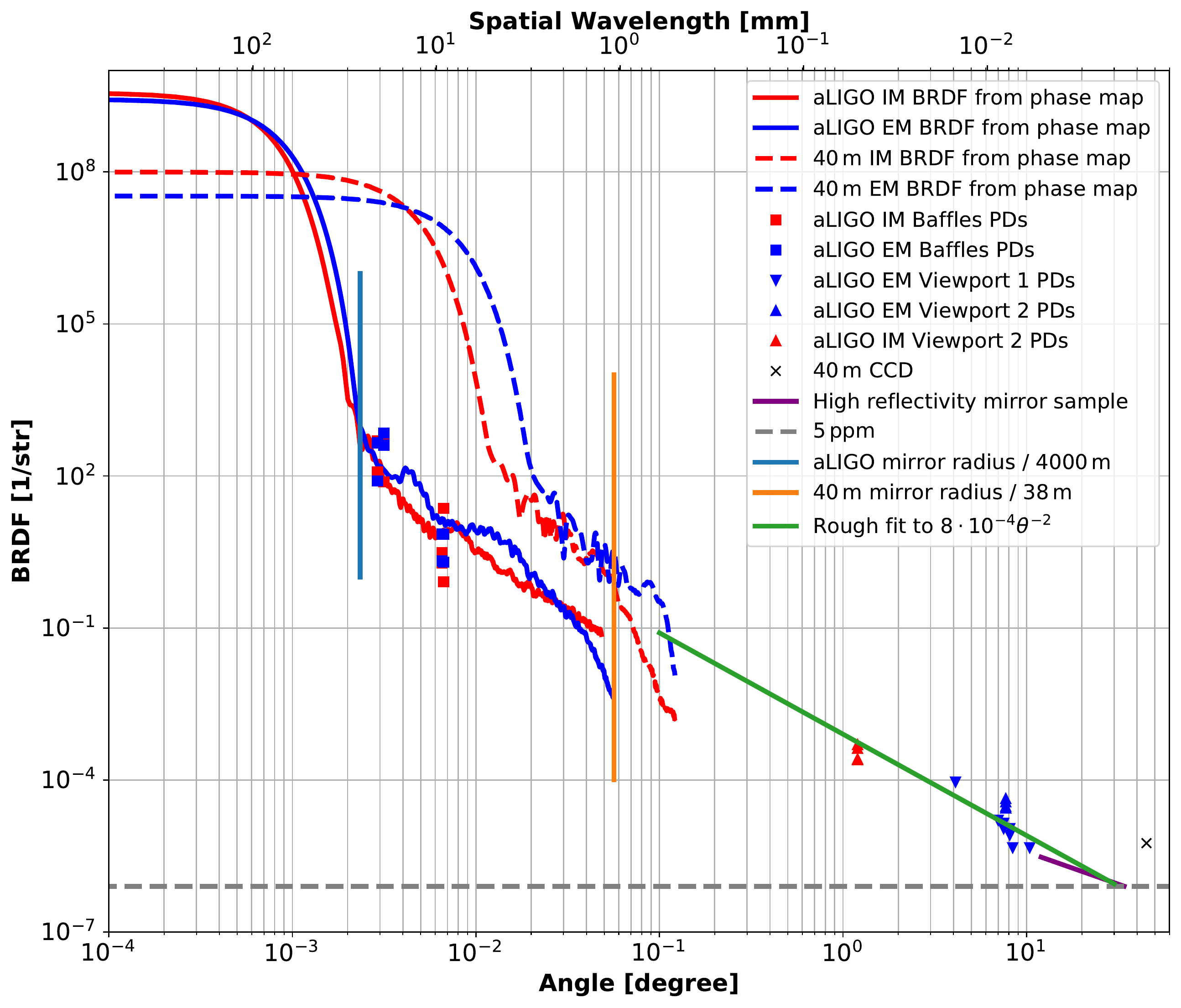}
		\caption{BRDF by various methods combined. Data colored in blue belongs to scatter by the EM (evaluated at the IM for small angles), and red by the IM (again, evaluated at the respective EM for small angles). The traces ending with 'from phase map' are BRDF estimates based on the phase maps of the LIGO Livingston Observatory (LLO) Y-arm and the 40$\,$m prototype interferometer mirrors (cf. \Cref{sec:phasemaps}). The cutoff angles at which scattered light completely misses the other mirror are also marked for the aLIGO (blue) and 40$\,$m (orange) designs. The fitted line for the data out of the end mirror has an angular dependence of $~\theta^{-2}$. Measurements of scattered light by photodiodes located at various points along the LIGO arms are shown in boxes and triangles. PDs located on the baffles surrounding the arm cavities detect light scattered at low angles while PDs located at the view ports near the aLIGO mirrors detect light at high angles. The single scattering measurement done at the 40\,m prototype interferometer is shown as x (cf. \Cref{sec:CCD}). The BRDF behavior measured for superpolished mirrors at high angles is marked by a purple line together with a 5\,ppm level in dashed grey indicating the constant BRDF behavior at high angles \cite{Magana-Sandoval_2012_Large-Angle-Scatter}.}
		\label{fig:BRDFdata}
\end{figure}

\Cref{fig:BRDFdata} summarizes various BRDF measurements and illustrates this gap. The solid and dashed line plots for aLIGO and the 40\,m show a sharp Gaussian towards $\theta \sim 0$, whose width is determined by the Gaussian beam size resonating in the cavity, and a tail structure based on the phasemap data that is indicative of approximate power law scaling. Note that in all cases, the BRDF line plots cover the entire angular range spanned by the mirror at the other end of the cavity, marked by respective vertical lines in \Cref{fig:BRDFdata}. For the aLIGO detectors, we have also added BRDF measurements at $\theta \simeq 0.003^{\circ}$ and $\theta \simeq 0.006^{\circ}$, taken from four photodetectors (PDs) on baffles that are located 20\,--\,40\,cm from the center of the mirror~\cite{SURF_Jia2018_LLO-loss}. As discussed in this reference, the power distribution is very sensitive to the PD location relative to the center of the mirror due to the mirror aperture clipping effect, and the measured powers vary by an order of magnitude, but is consistent with the estimate based on phasemap data.

The region outside of the mirror corresponding to $\theta > \theta_{cut}$ (cf. \Cref{tab:FPparams}), a power law fit to the data points yields an angular dependence of $\theta^{-n}$ with $n \sim 2$. Note that this loss estimation is much larger than the typical estimation using the Gaussian beam shape, with a contribution also coming from the clipping effect of the mirror. We also have sparse BRDF measurement data in the range of $1^{\circ}$ to $10^{\circ}$ collected from various view ports located along the beam tubes. Moreover, BRDF measurement of a high reflecting mirror in the angular range larger than $10^{\circ}$ degrees was reported in~\cite{Magana-Sandoval_2012_Large-Angle-Scatter}. This BRDF data and the aLIGO BRDF data using view ports connect smoothly. 

Total integrated scattering (TIS) from the mirror surface in some angle range can be calculated from the BRDF as follows:
\begin{equation}
    \label{eq:TIS}
    \text{TIS}=2\pi\int_{\theta_1}^{\theta_2}\text{BRDF}(\theta)\cos(\theta)\sin(\theta)d\theta
\end{equation}
The integral of the fitted line in \Cref{fig:BRDFdata} between $1^{\circ}$ and $75^{\circ}$ is 8\,ppm. This angular range corresponds to the range captured by seperate TIS measurements done on our mirrors. More details can be found in \Cref{s:TIS}.

\section{Measurement Techniques}
\label{s:Measurements}
\begin{figure}[tbp]
	\centering\includegraphics[width=\columnwidth]{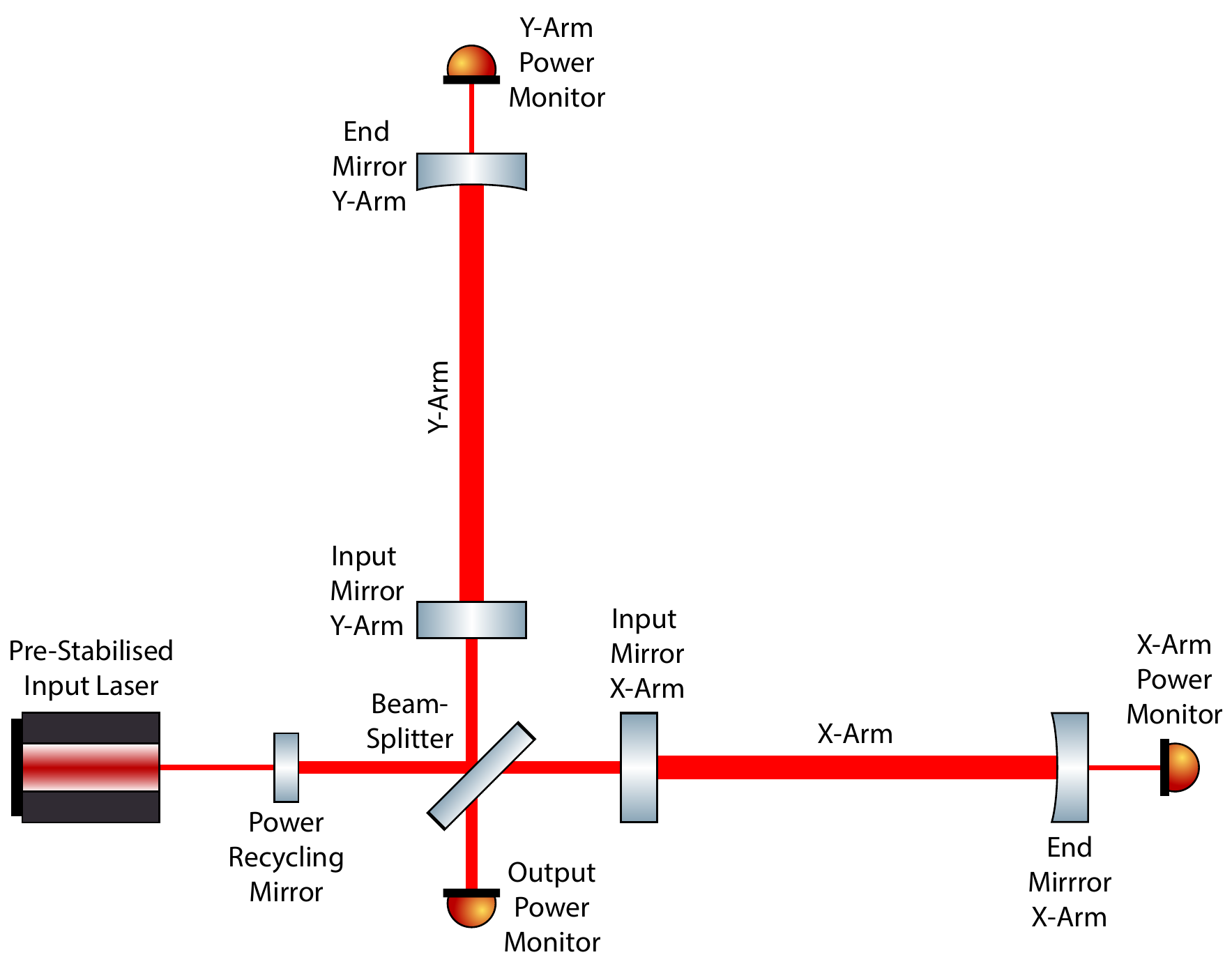}
	\caption{Optical configuration of the Caltech 40\,m prototype interferometer. The pre-stabilized laser is sent into the Fabry Perot Michelson interferometer and measured at the output mode monitor. The power recycling mirror (PRM) forms the impedance-matched power recycling cavity (PRC) with the rest of the interferometer. The arm cavities, formed by IMs  and EMs  increase the interaction time between stored photons and passing GWs in the full-scale instruments. The power in the arm cavities is monitored by PDs measuring the light transmission through the EMs.
	}
	\label{fig:40m-config}
\end{figure}

The Caltech 40\,m interferometer, 40\,m for short, is a small-scale GW detector prototype instrument. It was designed to share its optical topology with the full-scale LIGO observatories, including FP cavity-enhanced Michelson arms, a power recycling cavity (PRC), a signal recycling cavity (SRC), and an input mode cleaner (IMC). The optical configuration of the 40\,m is shown in simplified manner in \Cref{fig:40m-config}, which shows only the principle building blocks. 
Among similar instruments, the 40\,m's long arms and the comparatively large beam spot sizes on its input and end mirrors make the 40\,m an important subject for investigating optical scatter. Its optics were produced to the same standard as full-size GW detector mirrors, bringing its optical configuration closer to LIGO than any other prototype instrument.

We investigate optical losses in the FP arm cavities of the 40\,m to bridge the gap between modelled and measured optical scatter. For this we use a variety of complementary techniques, including:
\begin{enumerate}[label=(\roman*)]
\item Direct scattered light measurement by CCD
\item Profile maps of the mirror surfaces combined with wave propagation simulations (phase maps)
\item Total Integrated Scatter (TIS) measurements as a function of position on the mirror surface
\item Static optical cavity impedance measurement
\item Power recycling gain constraints
\end{enumerate}
In the following section we describe how these measurements are performed, what their systematic errors are, and determine if they can provide a complete picture of the scatter loss distribution.

\subsection{Direct scattered light measurement with cameras} 
\label{sec:CCD}
An in-situ measurement of the total integrated scatter at the Caltech 40\,m prototype is not possible due to the geometry of the vacuum envelope. However, the arm cavity beam spots on the mirrors are visible from a small number of viewports with a direct line of sight. These allow a direct assessment of the scatter amplitude at certain angles from the beam axis in particular directions, determined by the relative positions and orientations of the mirrors and viewports.

We performed a proof or principle measurement to obtain an independent estimate of the intra-cavity scatter using this approach. For this, we calibrated the pixel scale of a standard CCD Gig-E camera at relevant combinations of sensor gain and exposure time settings against a precision optical power meter. The light source for the calibration process was an power-monitored diode laser with the same wavelength as the main laser.
The calibration allows us to calculate the total optical power captured by the CCD sensor from its cumulative pixel values.
\begin{figure}[tbp]
	\includegraphics[width=\columnwidth]{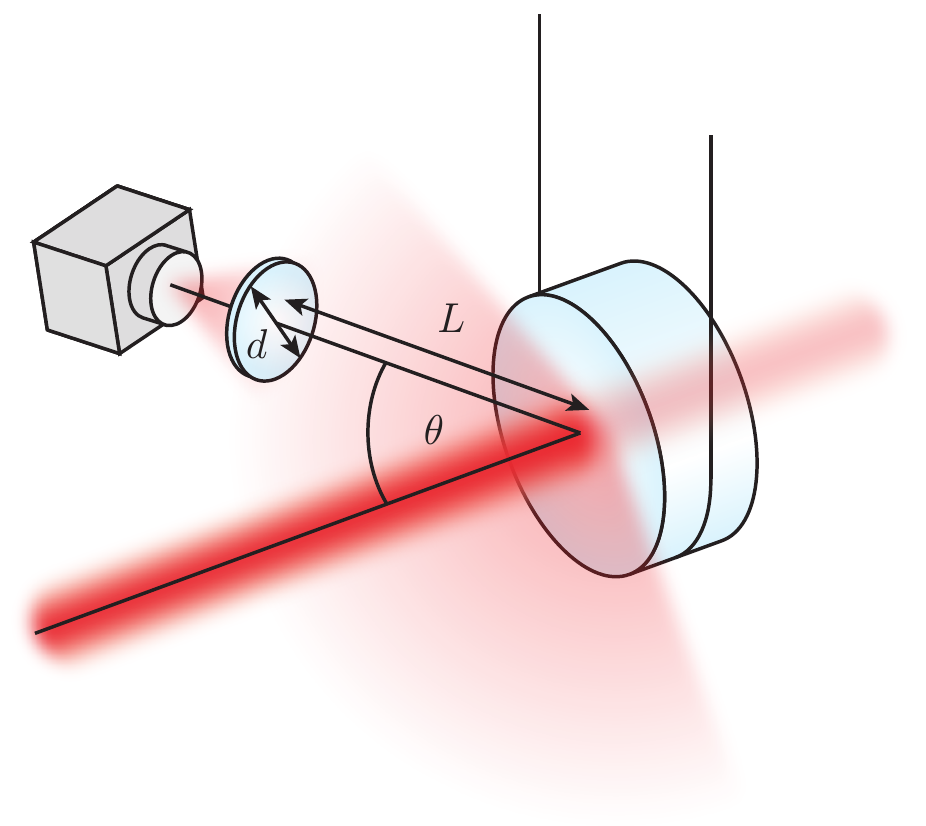}
	\caption{Evaluating large angle scatter of a suspended mirror with a CCD camera. Setup used to image large angle scatter ($\theta\sim$~\SI{50}{\degree}) from the beam spot on the mirror into a CCD camera using a $d = 5$\,cm diameter lens. The distance from the image to the lens is $L = $ \SI{65}{\centi\meter}.}
	\label{fig:ccd-setup}
\end{figure}

At the time of this study, only a single viewport on the vacuum chamber housing the EM on the X arm (EMX) was practical and available for this investigation. The given setting allowed us to place a focusing lens at a distance of 65\,cm at an angle close to $50^\circ$. The lens has a standard AR-coating, such that power loss to reflections can be neglected, and a clear aperture of 2 inches for photon collection. The arrangement is schematically shown in \Cref{fig:ccd-setup}. The lens' focal length and distance from the camera was chosen such that it images the beam spot on the mirror onto the CCD chip. We spatially filter the recorded CCD image with a digital aperture to reject stray light from parasitic light sources.

In addition to spatial image filtering and to isolate the pixel readings of the beam spot and remove all traces of ambient light scatter, we repeatedly turned the input laser feed into the interferometer on and off with a mechanical shutter and recorded exposures of the mirror with and without beam illumination. Subsequent exposures were thus taken in similar conditions, with the presence of the beam spot being the only difference, allowing us to use image subtraction to isolate the scattered power.

In total we measure the power scattered onto the CCD to be \SI{193.9 \pm 2.2}{\nano\watt} in the non-power recycled state with misaligned PRM. We use this measurement to estimate the bidirectional reflectance distribution function (BRDF discussed in \Cref{sec:BRDF}). Comparing this scattered power to \SI{10}{\watt} of circulating power in the arm cavity -- witnessed by the EM transmission -- we calculate the BRDF of the 40\,m's EMX at \SI{50}{\degree} to be \SI{6.2 \pm 1}{ppm\per\steradian}. This estimate was included in the compilation of BRDF measurements in \Cref{fig:BRDFdata}. 
 
The uncertainty in the BRDF result stems mainly from the uncertainty in the exact distance between beam spot to the focusing lens, which is estimated to be \SI{5}{\centi\meter}.

\subsection{Phase map measurements}\label{sec:phasemaps}
A phase-shifting Fizeau inteferometer (PSFI) \cite{zygo} was used to measure the surface figure of our mirrors. In such optical profilometers, a laser beam reflected from the sample under test is interfered with a beam reflected from a reference flat. A phasemap is then constructed by extracting the optical phase difference between the beams for every point on the surface from the interference pattern.

Phase maps for the 40\,m mirrors (IM4, EM7) were recorded at the time of installation with no significant qualitative differences between the different optics. However, while indicating that phase maps of different optics can be used interchangeably for simulation purposes, these initial phase maps were found to not be accurate enough for the estimation of the arm loss, especially in the short spatial wavelength region. 

To extend the spatial frequency range of our phase maps, new measurements were done using spare mirrors (EM6, IM1, IM2), which were polished and coated as part of the same batch as the installed optics by the same companies for each process. Unfortunately, the coated surface profiles in the long wavelength region do show differences between mirrors, even though all were produced with the same polishing and coating process.

Since the currently installed mirrors cannot be remeasured in a PSFI without removing them from their suspension cages in the vacuum envelope, we synthesize their phase maps by combining low spatial frequency data from the original measurement with high spatial frequency data from the spare mirrors. This was done by blending the lowpass filtered original maps with high pass filtered new maps at a cross-over wavelength of 1.3\,mm.

\begin{figure}[tbp]
\centering
		\subfloat[][]{\includegraphics[width=0.5\columnwidth]{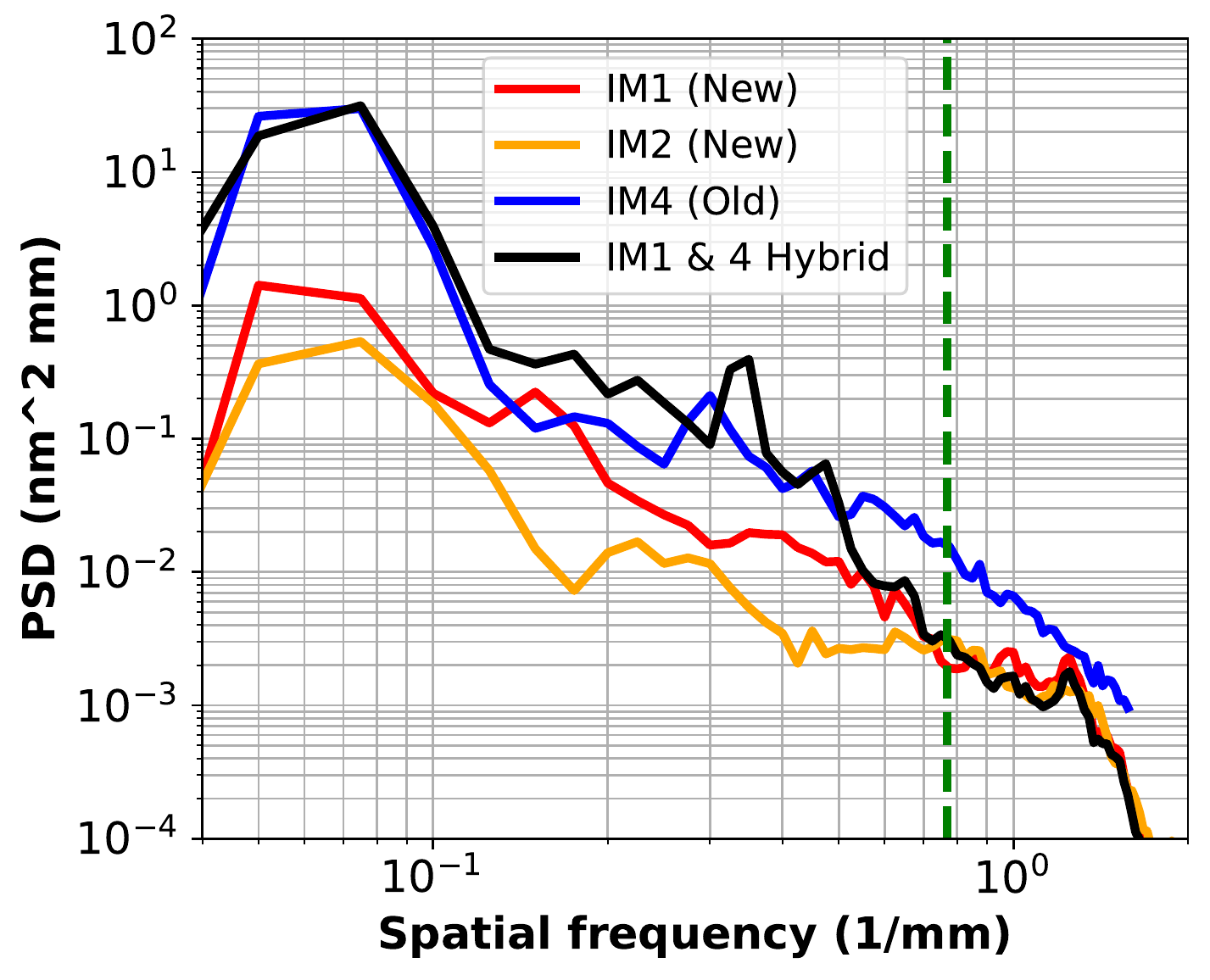} \label{fig:ITM40mPSDs.pdf}}
		\subfloat[][]{\includegraphics[width=0.5\columnwidth]{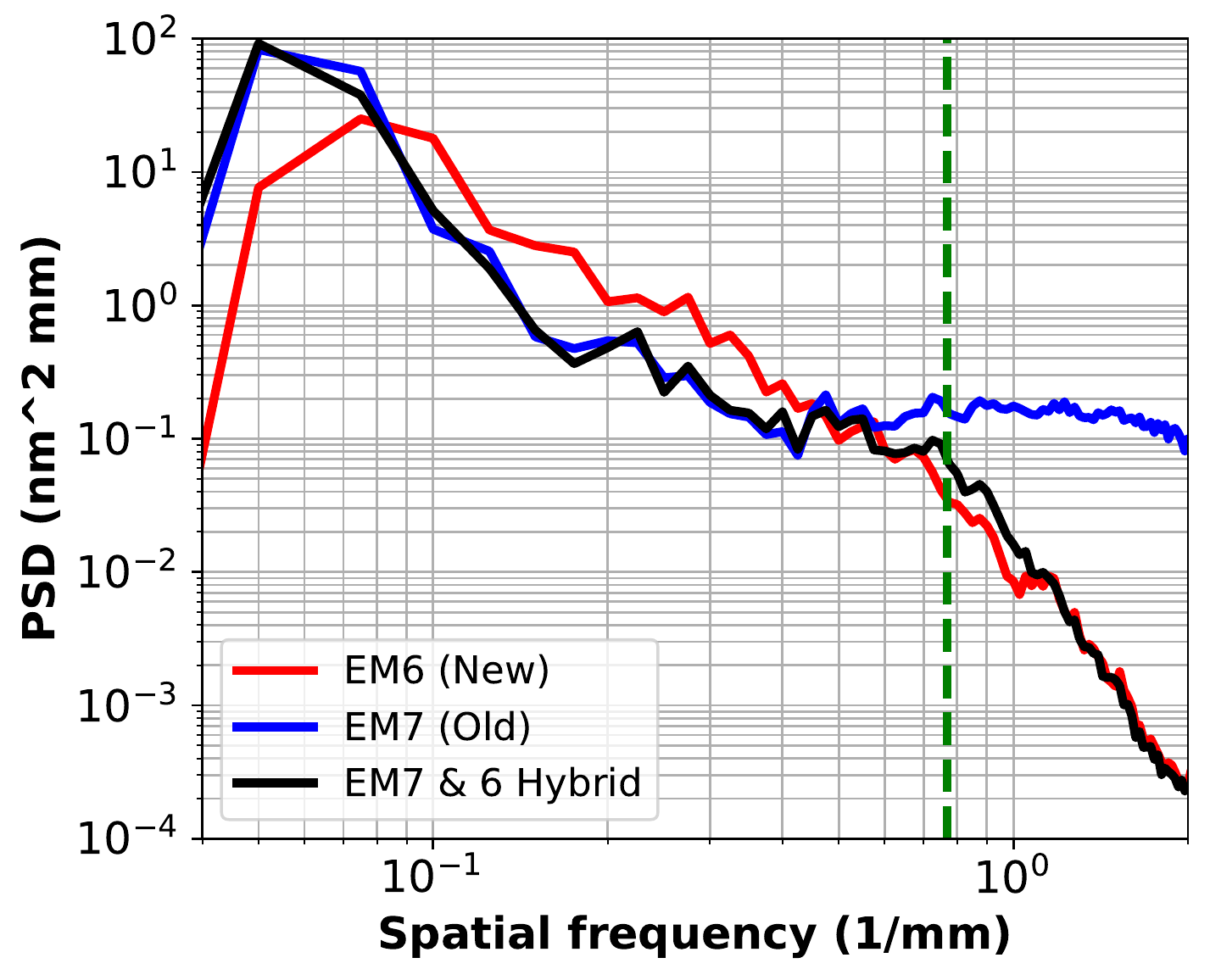}\label{fig:ETM40mPSDs.pdf}}
		
		\subfloat[][]{\includegraphics[width=0.5\columnwidth]{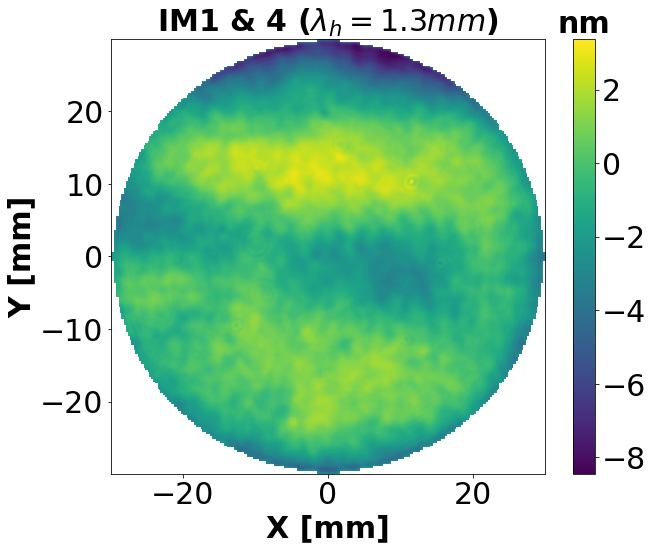}\label{fig:ITM1_4Hybrid_Phasemap.png}}
		\,\,\,\,\subfloat[][]{\includegraphics[width=0.5\columnwidth]{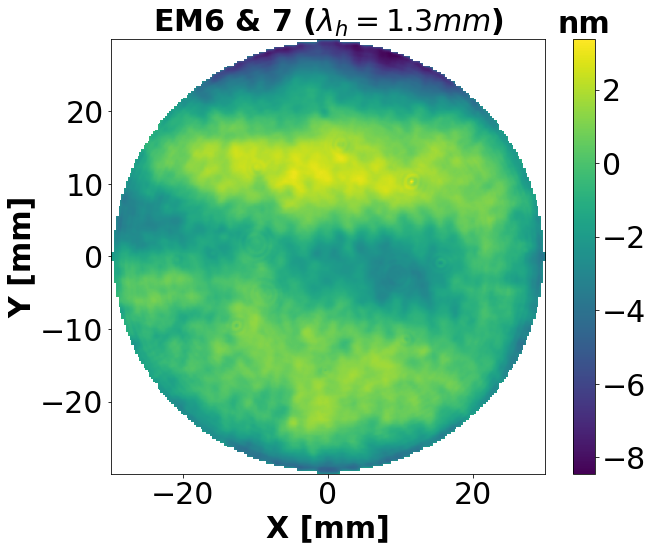}\label{fig:ETM6_7Hybrid_Phasemap.png}}
		\caption{PSDs of (a) IMs. (b) EMs. Green dashed lines mark the spatial frequency corresponding to $\lambda=\SI{1.3}{\milli\meter}$, where the old and new maps were hybridized. Hybrid phase maps of (c) IM1 and IM4 (d) EM6 and EM7.}

\end{figure}

In similar fashion to \Cref{eq:BRDF_1D}, 1D PSDs were computed from the 2D PSDs of the phase maps by averaging over the azimuthal angle. The PSDs of the newly measured phase maps of spare IMs - IM1 and IM2 are shown in \Cref{fig:ITM40mPSDs.pdf}. These maps were measured focusing in the short wavelength region, $\lambda<\SI{5}{\milli\meter}$. The comparison of the red (IM1) and orange (IM2) lines show that the PSDs in $\lambda_s<\SI{1.5}{\milli\meter}$ are the same. From this result, the same PSD can be assumed for the IM4, which is the one actually installed. The black line is a combination of the ITM04 map (blue) original measurement with the IM1 new measurement. The cutoff for the hybridization is $\lambda=\SI{1.3}{\milli\meter}$ (green dashed line).

A corresponding analysis was done for the EMs. \Cref{fig:ETM40mPSDs.pdf} is the comparison of PSDs of old EM7 map (blue), new EM6 map (red) and the hybrid of these two (black) merged at $\lambda=\SI{1.3}{\milli\meter}$.

These hybrid phase maps are used in a numerical simulation to simulate beam propagation \cite{yamamoto2007sis} in a cavity. In the simulation, an input Gaussian laser beam, mode matched to the cavity, is bounced back and forth by the cavity mirrors loaded with phase maps until a steady-state is reached. The beam propagation between the cavity mirrors is computed using Fourier optics methods. At steady state, the roundtrip loss is computed by integrating over the intensity of the beam outside the mirrors' apertures. For the case of a 40\,m cavity with mirrors loaded with the hybrid phase maps shown in \Cref{fig:ETM6_7Hybrid_Phasemap.png} and \Cref{fig:ITM1_4Hybrid_Phasemap.png}, the roundtrip loss was computed to be $\SI{6 \pm 2}{ppm}$ over the angular range accessible by this method. 

\subsection{Total Integrating Scatter Measurement } \label{sec:TIS}
\label{s:TIS}
\begin{figure}[htbp]
\centering
      \includegraphics[width=\columnwidth]{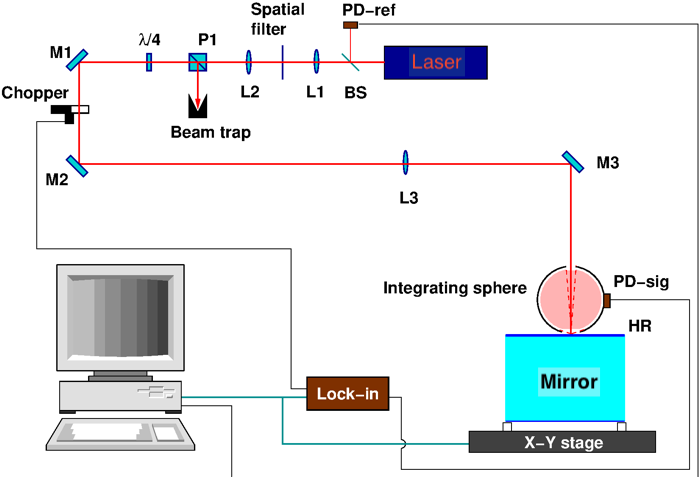}
      \caption{A schematic of the measurement setup of total integrated scattering. Details of the setup are described in the main text. (PD) Photodiode (BS) Beamsplitter (L) Lens (P) Polarizer ($\lambda/4$) Quarter wave-plate (M) Mirror (HR) High reflection side.
      }
      \label{fig:TISFig1}
\end{figure}

The total integrated scattering (TIS) from a HR coating is measured using an integrating sphere. \Cref{fig:TISFig1} shows a schematic of the experimental setup. 

The 1064 nm output of a JDSU NPRO laser is first sampled for a reference photodiode (PD-ref) and spatially filtered to generate a high-purity TEM00 mode. A sequence of polarizer, quarter wave-plate and chopper constrains its polarization and allows for lock-in detection. The collimated beam enters the top port of the integrating sphere and a mode-matching lens is used to position the beam waist at the mirror's HR surface. The reflected specular beam is aligned to exit the integrating sphere centered on the top port. The mirror is raster scanned and the scatter from HR mirror is measured by a signal photodiode (PD-sig) attached to the integrating sphere to obtain the TIS map. The beam diameter and the scanning step size are both $\SI{300}{\micro\meter}$. The polar angle of scatter collection of the integrating sphere is defined by the sizes of the top and bottom ports, the current set-up has a collection range from $\SI{1.0}{\degree}$ to $\SI{75}{\degree}$, corresponding to a bandwidth of spatial frequency $16$ to $\SI{908}{\milli\meter^{-1}}$. The calibration is carried out by placing a diffuse reflectance standard (Labsphere Spectralon) at the HR mirror position. 

\begin{figure}[htbp]
	\centering
	\subfloat[][]{\includegraphics[width=0.5\columnwidth]{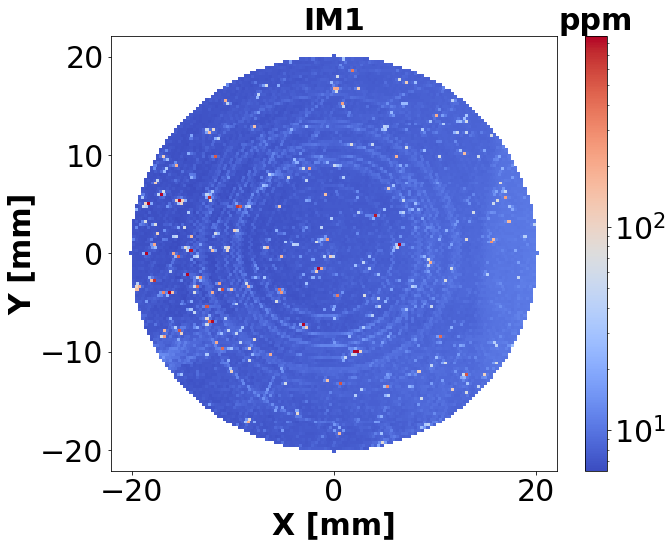} \label{fig:TIS_ITM_Sample1}} \subfloat[][]{\includegraphics[width=0.5\columnwidth]{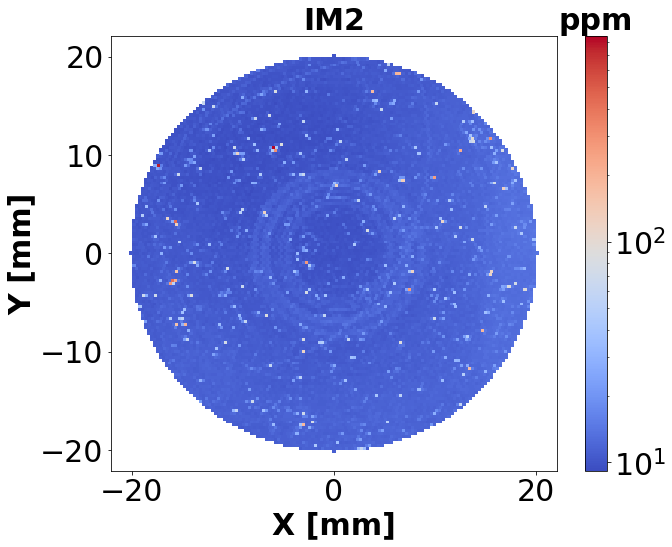} \label{fig:TIS_ITM_Sample2}}
	
	\subfloat[][]{\includegraphics[width=0.5\columnwidth]{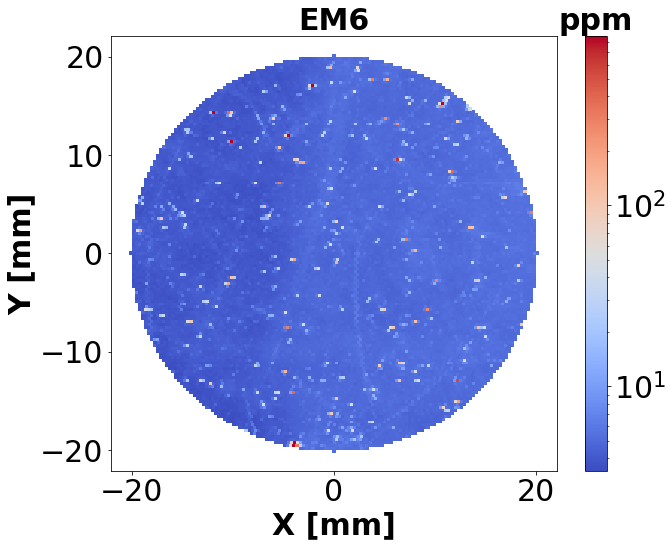} \label{fig:TIS_ETM_Sample1}} \subfloat[][]{\includegraphics[width=0.5\columnwidth]{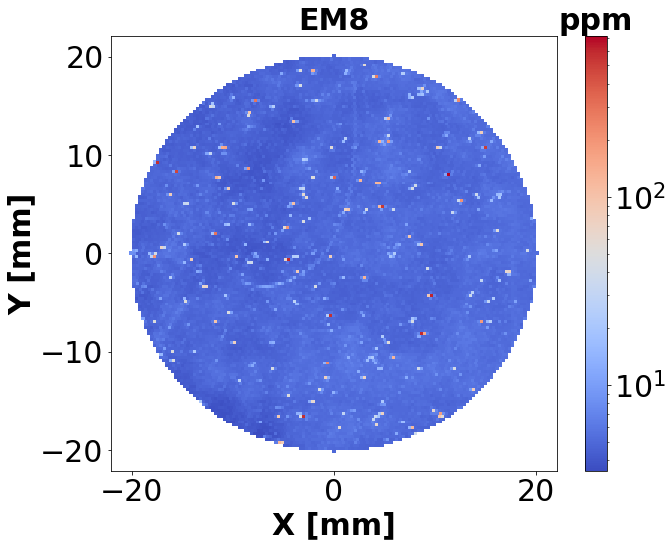} \label{fig:TIS_ETM_Sample2}}		
	\caption{Reflected power fraction measured by a laser for (a) IM1 and (b) IM2. (c) EM6 (d) EM8
	}
	\label{fig:TISSample}
\end{figure}
\begin{figure}		
	\subfloat[][]{\includegraphics[width=0.5\columnwidth]{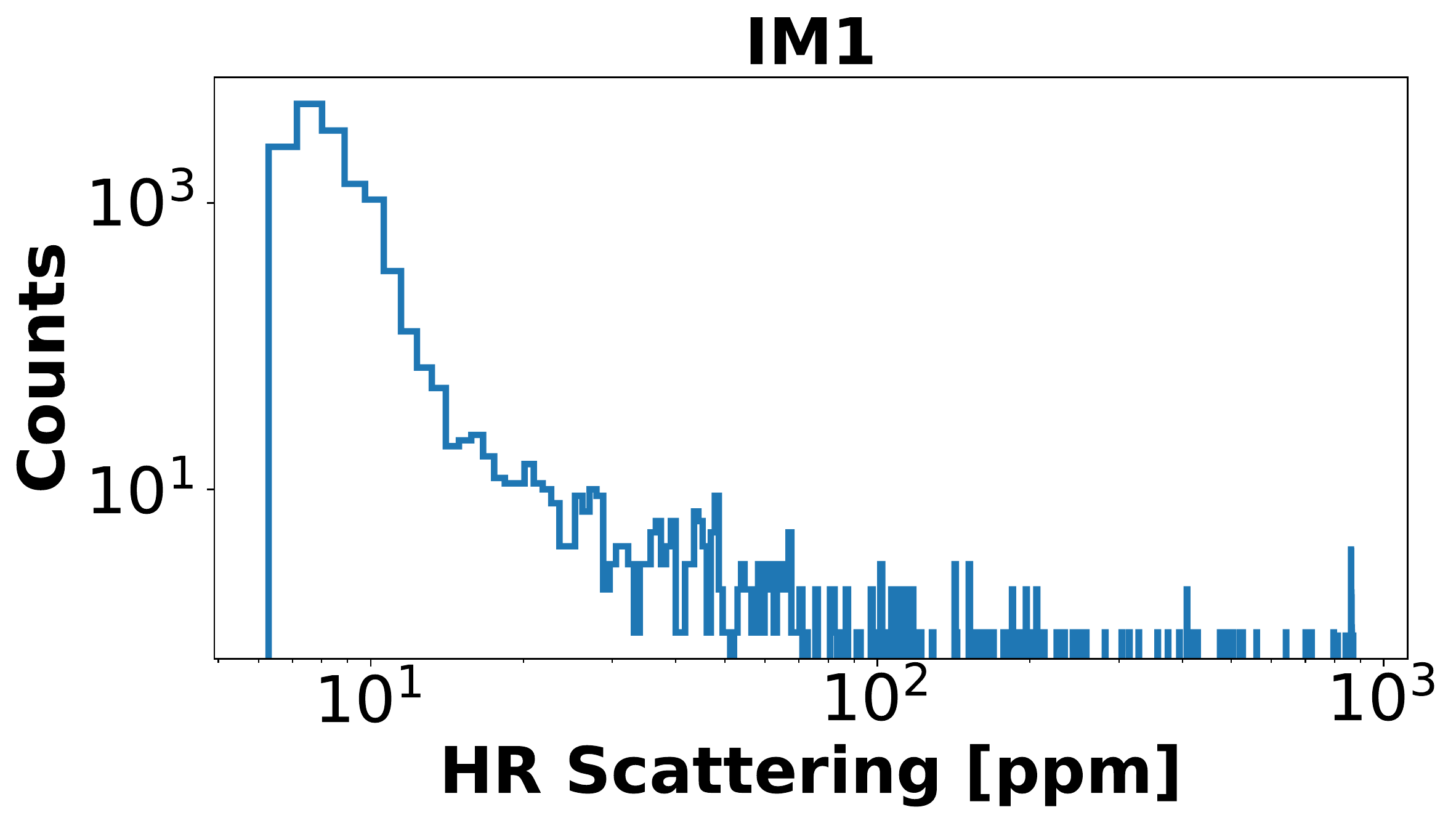}} \subfloat[][]{\includegraphics[width=0.5\columnwidth]{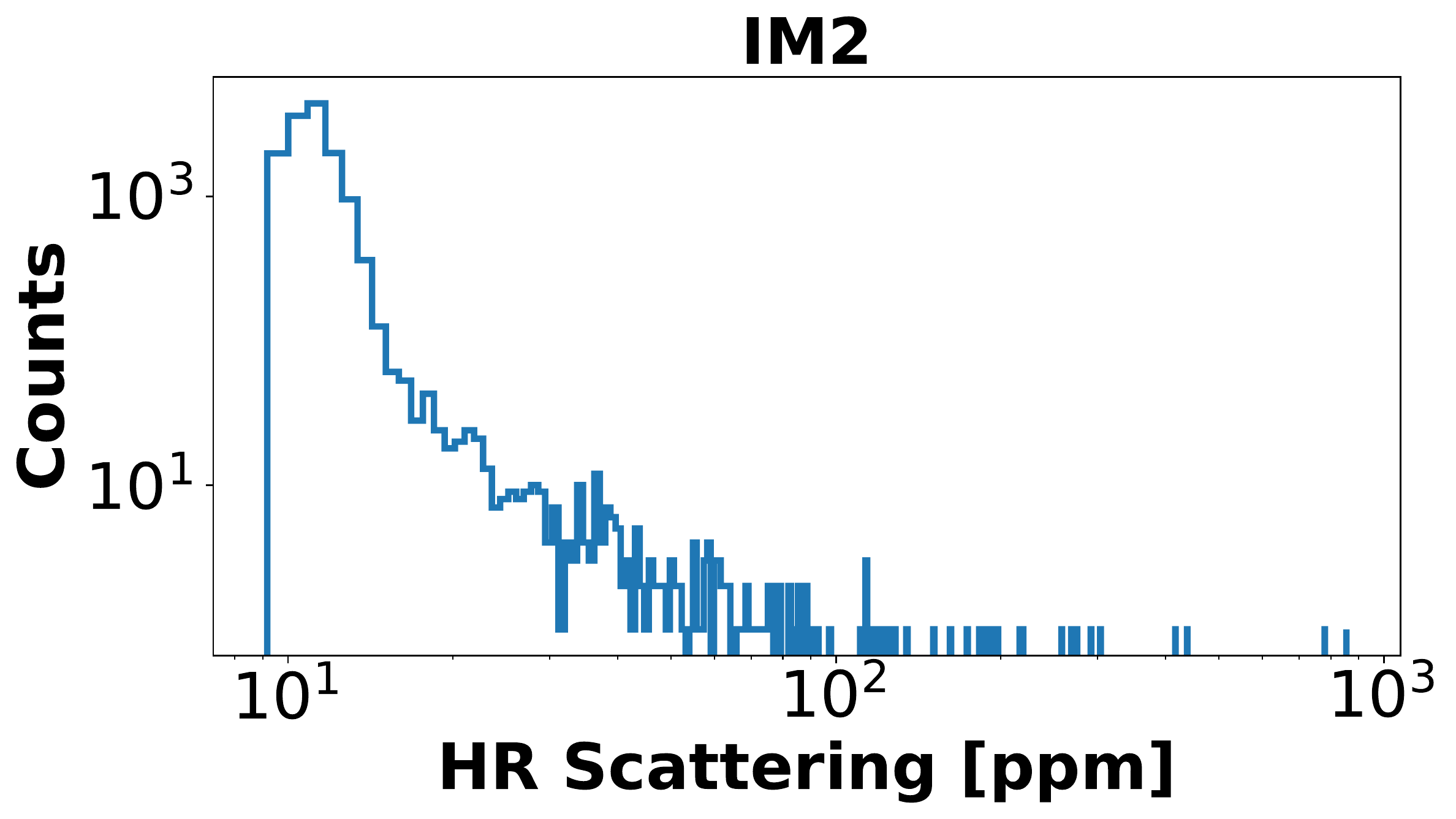}}
	
	\subfloat[][]{\includegraphics[width=0.5\columnwidth]{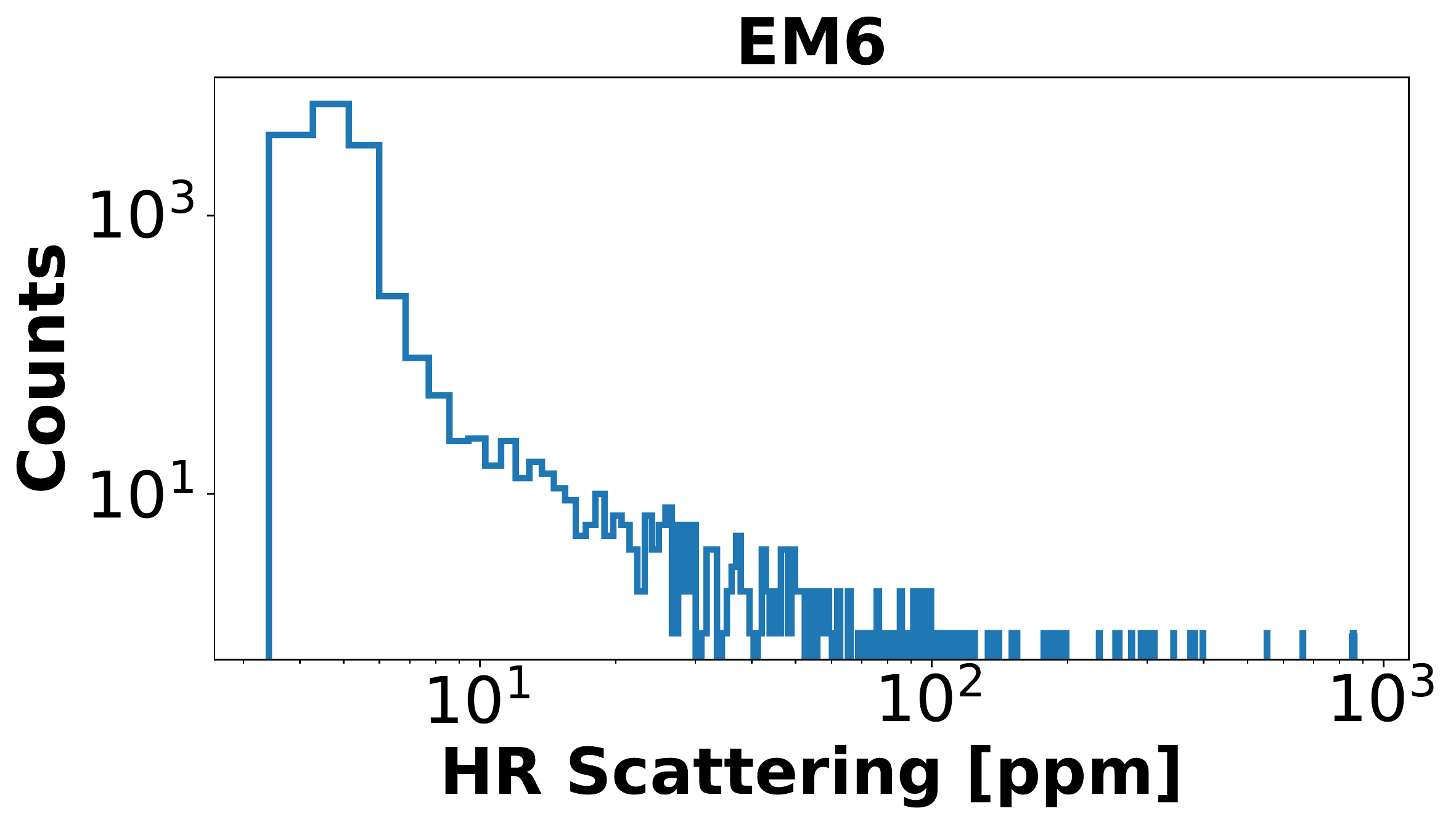}} \subfloat[][]{\includegraphics[width=0.5\columnwidth]{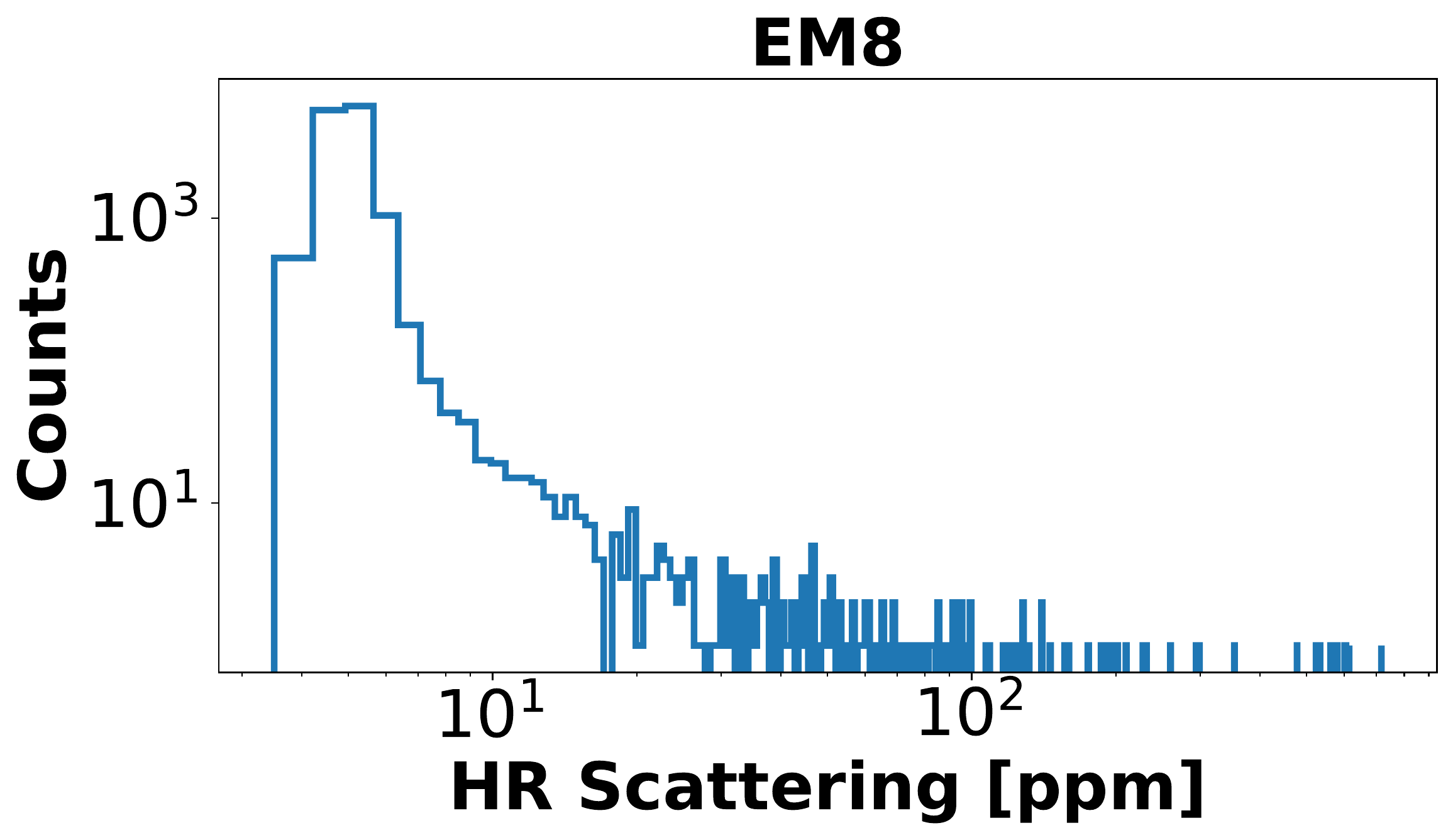}}

	\caption{Populations of scattered power fraction in (a) \Cref{fig:TIS_ITM_Sample1}, (b) \Cref{fig:TIS_ITM_Sample2}, (c) \Cref{fig:TIS_ETM_Sample1} and (d) \Cref{fig:TIS_ETM_Sample2}.
	}
	\label{fig:TISHist}
\end{figure}

\Cref{fig:TISSample} shows TIS measurements that were pereformed on four 40\,m prototype interferometer spare mirrors (IM1, IM2, EM6, EM8). \Cref{fig:TISHist} shows the histograms of the power fractions in \Cref{fig:TISSample}. The histograms are peaked at $\sim$7\,ppm for the IMs and $\sim$3\,ppm for the EMs, and there is a long tail extending over 1000\,ppm.

We calculate the total scatter loss in the arm cavities to angles between $\SI{1.0}{\degree}$ and $\SI{75}{\degree}$ in the following way. A Gaussian beam, the same size as in the 40\,m arm cavities, is impinging on the center of the mirror. Each point in the TIS map induces a scatter loss to the part of the Gaussian beam that overlaps with it. The losses are then summed incoherently to obtain the total scatter loss. Since we are summing over large angle range interference effects are heavily suppressed.

Explicitly, the total scatter loss is given by
\begin{equation}
	\text{Loss} = \frac{1}{P_0}\sum_n \text{TIS}[x_n,y_n] |E_q[x_n,y_n]|^2\pi\omega_\text{TIS}^2,
\end{equation}
where $ \text{TIS}[x_n,y_n]$ are the TIS measurements data points, $E_q[x_n,y_n]$ is the normalized electric field distribution of a Gaussian beam the size of the beam impinging on that mirror and $\omega_\text{TIS}$ is the TIS beam waist. Doing this summation for each sample we find that the scatter losses, to angles between $\SI{1.0}{\degree}$ to $\SI{75}{\degree}$, are \SI{10\pm 1.7}{ppm},  \SI{8.6\pm 0.4},  \SI{4.6\pm 0.03} and  \SI{4.9\pm 0.14} in IM1, IM2, EM6 and EM8, respectively. The errors are calculated by shifting the beams randomly around the centers of the mirrors with an standard deviation of \SI{1}{\milli \meter} and computing the standard deviation of the resulting losses.

\subsection{Static Optical Impedance Measurement} \label{sec:DCMethod}
Inferring cavity round-trip losses from measuring either optical ringdown time constants or the linewidths of cavity resonances and accounting for the known mirror transmissivities has a significant caveat: The full uncertainty in the measured mirror transmissivities is directly transferred to the loss estimate. For GW detector arm cavities, which have moderately high finesse and are highly over-coupled, the uncertainty of the IM transmission is usually on the same order or larger than the expected losses.

In a measurement of the optical cavity impedance, also referred to as the \textit{DC-Method}, we compare the static reflected power from the IM in two cases: the prompt reflection off the IM when there is no arm cavity (either because the EM is misaligned or because the intra-cavity beam path is blocked), and when the arm cavity is on resonance and locked to the laser carrier. The combination of the two individual measurements allows to solve for the cavity losses with reduced sensitivity to the uncertainty in IM transmission. This method only measures total optical loss in the cavity and cannot distinguish absorption from scatter, however, the optical absorption of the high-quality tantala-silica coatings used in GW detectors and prototypes is of order \SI{1}{ppm} or smaller and can be safely neglected.

It carries similarities to the optical ringdown measurements that are carried out in reflection described by Isogai et al.~\cite{Isogai_2013_LossStorageTime}, but is more suited to the optical topology of a cavity-enhanced Michelson interferometer, in which the IMC acts as a lowpass filter for any modulations of the input laser. All modulators and shutters are located upstream from the IMC, such that the fast switching of the input power that is required for proper ringdown measurements would get filtered by the IMC's optical response. The IMC's storage time is longer than those of the arm cavities, such that the measured ringdown time constants would be entirely dominated by the IMC, significantly limiting the achievable accuracy on arm cavity effect estimation.

We consider the cavity reflection coefficient on resonance $R_r$ and off resonance $R_m$ and the coupling inefficiency of input beam into the cavity. We find the expression for the ratio of the two reflected power measurements $\frac{P_r}{P_m}$ to be 
\begin{equation}
	\frac{P_r}{P_m}=\frac{1-\eta\gamma(1-R_r)}{R_m},
	\label{eq:PowerRatioVsLoss}
\end{equation}

$\eta$ is the mode matching efficiency of input beam to the arm, $\gamma$ is the fraction of power in the carrier frequency after sideband modulation. The values for these parameters are given in \Cref{tab:40m-parameters}. The expressions for $R_r$ and $R_m$ can be found in \cite{Isogai_2013_LossStorageTime}. Assuming the loss contribution from the two cavity mirrors is the same we extract the roundtrip loss from \Cref{eq:PowerRatioVsLoss} numerically.

\begin{figure}[h!]
\centering\includegraphics[width=\columnwidth]{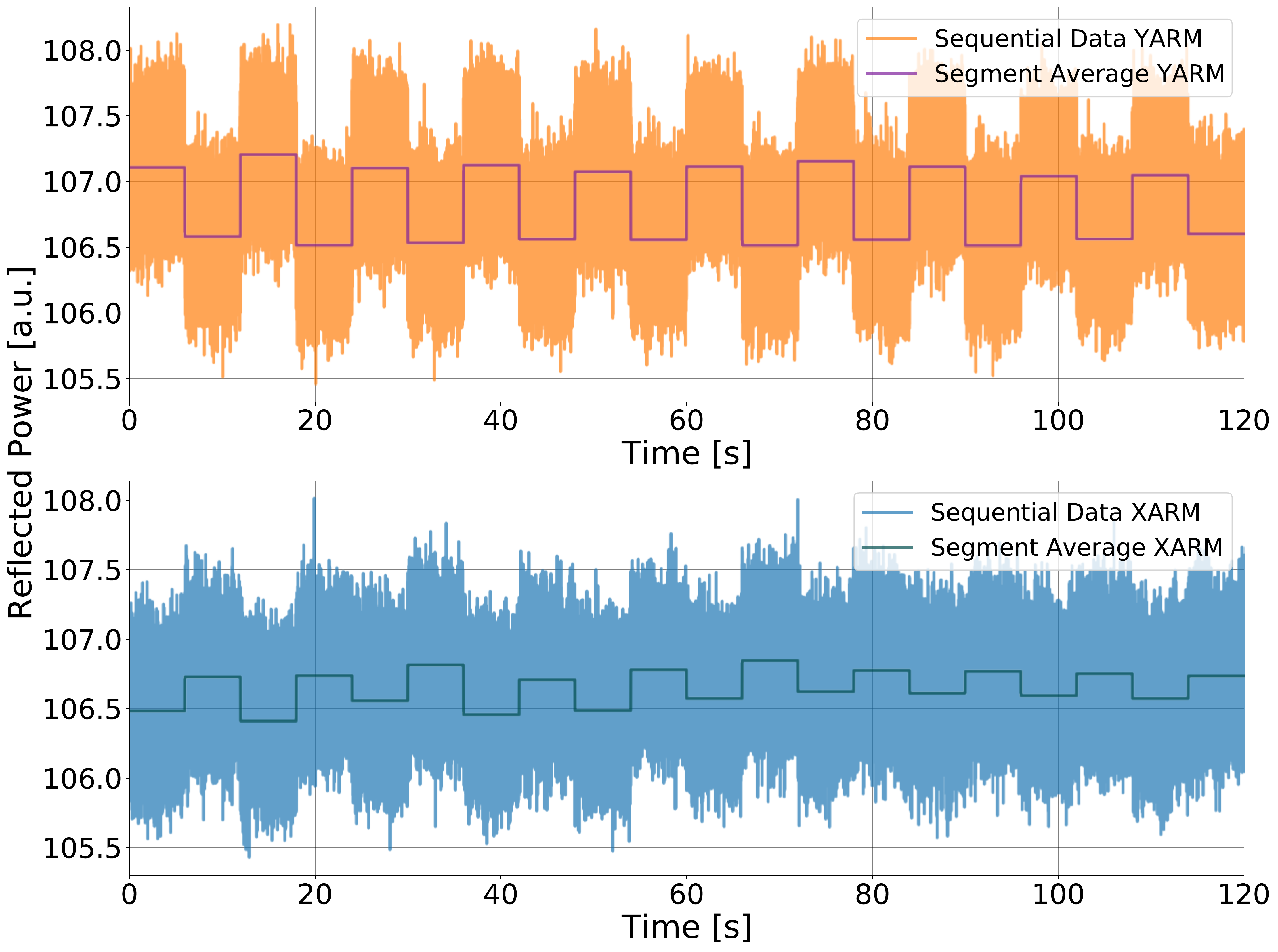}
\caption{DC reflection measurement sequence. The measured reflected power is shown as a function of time during which the measured cavity is locked and unlocked periodically. Top(bottom) figure shows the measurement sequence for the Y(X) arm of the 40\,m. The square wave lines show the average at each measurement segment. 
}
\label{fig:refl-seq}
\end{figure}

For very low roundtrip losses, such as in the case of the arm cavities in the 40\,m, the reflected power ratio gets very close to 1 which requires long integration times to get good enough resolution between the two power measurements. We repeat the arm lock/unlock cycle with re-alignment procedures in between to counteract drifts. During the measurement time we track the changes to the optical system (pointing/alignment drifts etc) and correct for fluctuations in the input power by scaling the reflected power to the circulating power in the IMC. An example of such a measurement sequence is shown in \Cref{fig:refl-seq}. The top (bottom) graph shows the measurement sequence for the Y(X) arm of the 40\,m. The two traces appear opposite in phase due to the fact that the X and Y arm losses are on opposite sides of the $\frac{P_r}{P_1}=1$ point. After initial measurements of the 40m arms, which showed contamination with excess losses, the mirrors were cleaned with First Contact solution. Our best estimates for the arm cavity losses post-cleaning are \SI{20 \pm 2.5}{ppm} for the X cavity and \SI{50 \pm 2.5}{ppm} for the Y cavity.

\begin{center}
	\begin{table}
		\renewcommand{\arraystretch}{2}
		\resizebox{\columnwidth}{!}{%
		\begin{tabular}{ @{\hspace{0.5cm}} c @{\hspace{0.5cm}}| @{\hspace{0.5cm}}c @{\hspace{0.5cm}}| @{\hspace{0.5cm}}c@{\hspace{0.5cm}} }
			\hline\hline
			Parameter &Nominal Value & Uncertainty\\ \hline
			
			IM Transmission (T1) & \SI{1.38}{\percent} & $\pm$ \SI{0.01}{\percent}\\
			
			EM Transmission (T2) & \SI{13.7}{ppm}& $\pm$ \SI{3}{ppm}\\
			
			Mode matching efficiency ($\eta$) & \SI{92}{\percent} & $\pm$ \SI{5}{\percent}\\
			
			Modulation depth @ \SI{11}{\mega\hertz}& 0.179 & \\
			
			Modulation depth @ \SI{55}{\mega\hertz}& 0.226 & \\
			
			Carrier frequency power fraction ($\gamma$) & \SI{95.92}{\percent} & \\
			\hline\hline
		\end{tabular}
		}
		\caption{40\,m intereferometer prototype arm cavity parameters used in \Cref{eq:PowerRatioVsLoss}.
		}
		\label{tab:40m-parameters}
	\end{table}
\end{center}

\subsection{Power Recycling Gain Measurement \label{sec:PowerRecycle}}

In power recycled interferometers, such as the 40\,m interferometer prototype, the recycling gain, the ratio between the power circulating the recycling cavity and the incident power, is inversely proportional to the round trip loss in the arm cavities. For the ideal case, where the power recycling cavity (PRC) is critically coupled, the power recycling gain $G$ is given by:
\begin{equation}
G = \frac{\pi}{2FL},
\end{equation}
where $F$ is the average arm cavity finesse and $L$ is the average round trip loss in the arm cavities. The systematic errors in this technique can be quite small since the loss influence on the PRG is amplified by the finesse of the arm cavities. In the more realistic case where the PRC is not exactly critically coupled, the expression for $G$ becomes more complicated, but the insensitivity to the uncertainty in the IM transmission remains.

To obtain an estimation of the PRG we measure the optical power transmitted through the arm cavities when the PRC is locked - $P_\text{Locked}$ and unlocked - $P_\text{Unlocked}$. The PRG is then
\begin{equation}
	G = T_\text{PRM}\frac{P_\text{Locked}}{P_\text{Unlocked}},
\end{equation}
where $T_\text{PRM}$ is the transmission through the power recycling mirror. 

One complication that arises is the uncertainty in the mode-matching between the PRC and the arm cavities which is hard to quantify. For the 40m, we know that the mode-matching between the input beam and the arm cavities, measured when the PRC is unlocked, is 92\% (\Cref{tab:40m-parameters}). In our calculations we assumed that the mode-matching between the input beam and the coupled cavitiy, that includes the PRC and the arm cavities, is the same. This assumption is consistent with the other loss and PRG measurements as can be seen below.

As mentioned before, the uncertainties in the PRG measurement can be quite small. Unfortunately, the optical loss in the 40\,m's PRC, which is required in the analysis, suffered a comparably large systematic uncertainty at the time of measurement. Due to coating stress, the folding mirrors of the 40\,m PRC have unintended marginally convex HR surfaces, which drives the original PRC design into the geometrically unstable regime. Flipping the folding mirrors solved this stability issue, but as a result, the back-side AR coatings and mirror substrates are now part of the optical path in the PRC, which adds extra losses that are not fully characterized. Especially the residual reflection of the AR coatings adds significantly to the PRC loss uncertainty. As a result, the uncertainty in the inferred arm cavity loss is quite big. Our best estimate for the arm cavity loss from the PRG measurement, after the mirrors were cleaned, is \SI{37 \pm 16}{ppm}, consistent with the average loss measured using the static optical impedance measurement.

\begin{figure}[h]
\centering\includegraphics[width=\columnwidth]{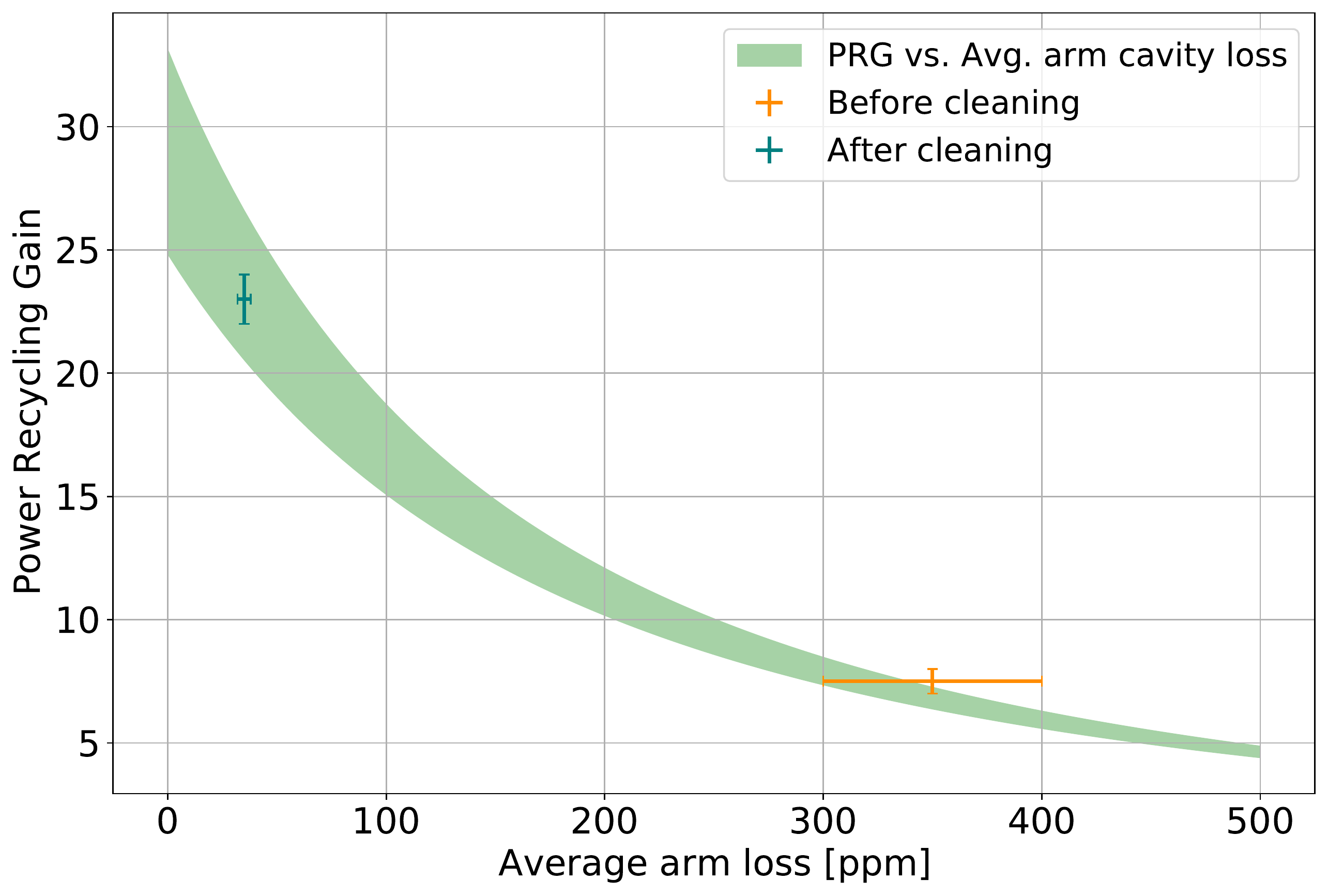}
\caption{Measuring loss using the PRG. The theoretical dependence of PRG on the average arm cavity loss is plotted as a green shade. The shade width is due to the big uncertainty in the PRC roundtrip. Orange and blue data points indicate the PRG and average arm cavity loss measured using the DC reflection method before and after cleaning of the mirrors in the cavities, respectively. The widths of the errorbars indicate the 1$\sigma$ uncertainty of these measurements.
}
\label{fig:prg_dc_refl}
\end{figure}

The viability of the method can still be demonstrated. \Cref{fig:prg_dc_refl} shows the theoretical dependence of PRG on the average arm cavity loss computed using a Finesse model \cite{FINESSE_2004} that includes the losses in the PRC, their uncertainty, and the non-ideal mode-matching between the PRC and the arm cavities. On top of that plot, the average arm cavity loss, measured using the DC reflection method, and the measured PRG are shown as data points. The measurements before and after the cleaning of the mirrors in the 40\,m arms using first-contact are colored in orange and blue respectively. The widths of the error bars indicate the 1$\sigma$ uncertainty of these measurements. It can be seen that the intersection of these measurements agree quite well with the theoretical line demonstrating the consistency of these measurement techniques.
 
\section{Comparison of Measurements}
\label{s:Comparison}
\begin{figure}[!h]
	\centering\includegraphics[width=\columnwidth]{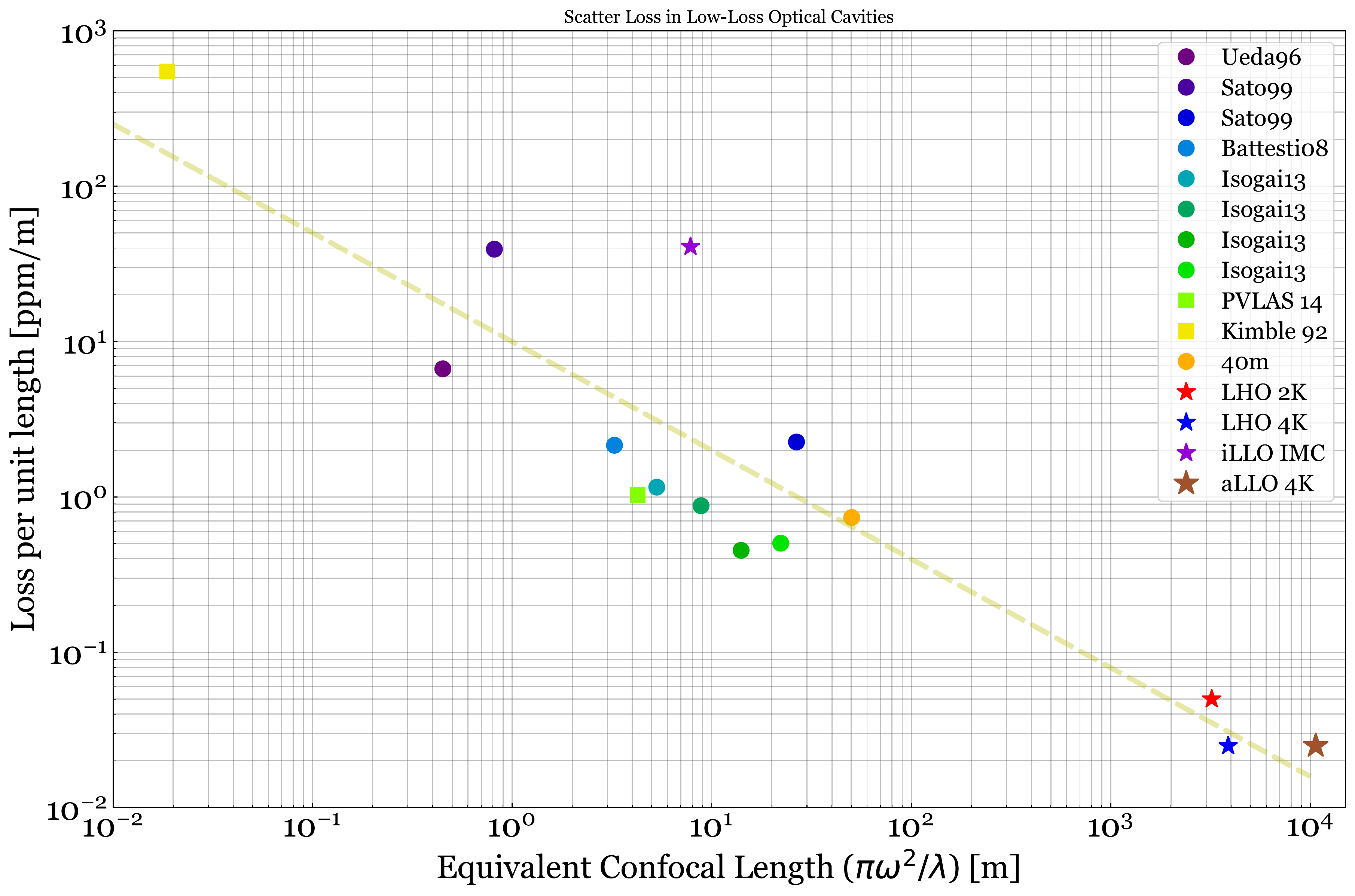}
	\caption{Loss per unit length as a function of equivalent confocal length. Our results (40\,m) are compared to previously published works: Ueda96 \cite{Ueda1996}. Sato99 \cite{Sato:99}. Battesti08 \cite{Battesti2008}. Isogai13 \cite{Isogai_2013_LossStorageTime}. PVLAS 14 \cite{DellaValle:14}. Kimble 92 \cite{Rempe:92}. The loss measurements done in the LIGO sites LHO 2K and LHO 4K measurements were done on the \SI{2}{\kilo\meter} and \SI{4}{\kilo\meter} arms at Hanford respectively during the initial LIGO phase. iLLO IMC is a measurement done of the IMC in the Livingston interferometer during the initial LIGO phase.  aLLO is a loss measurement done of the \SI{4}{\kilo\meter} arm at Livingston during the Advanced LIGO phase.
	}
	\label{fig:isogai}
\end{figure}

In the previous sections we discussed several direct (DC reflection, PRG) and indirect (mirror phase maps, total integrated scatter, scattering measurements) loss measurement techniques. The result of the comparison of these methods are summarised in \Cref{tab:paper_results}. The measured roundtrip loss using the DC reflection method after the mirrors were cleaned is quite low, making it suitable for comparison with the indirect methods. As described in \Cref{sec:PowerRecycle}, it is consistent with the PRG measurement.

\begin{table*}[t]
  \centering
  \renewcommand{\arraystretch}{1.5}
  \begin{tabular}{l|l|c |c |c} \hline \hline
  
	& Parameter & Nominal Value [ppm] & Uncertainty [ppm] & See\\ \hline
		\parbox[t]{4mm}{\multirow{2}{*}{\rotatebox[origin=c]{90}{Direct}}} 
		& Average static optical impedance measurement after cleaning & 35 & $\pm$ 3 & \Cref{sec:DCMethod}\\
		& PRG measurement after cleaning & 37& $\pm$ 16 &\Cref{sec:PowerRecycle}\\
		\hline
	\parbox[t]{4mm}{\multirow{6}{*}{\rotatebox[origin=c]{90}{Indirect}}} & Low-angle loss calculated from phase map& 6 & $\pm$ 2 & \Cref{sec:phasemaps}\\

	& IM Avg. TIS measurement ($\SI{1}{\degree}<\theta<\SI{75}{\degree})$ & 9.3 & $\pm$ 1.2 & \Cref{sec:TIS}\\
	& EM Avg. TIS measurement ($\SI{1}{\degree}<\theta<\SI{75}{\degree})$ & 4.8 &$\pm$ 0.2 &  \Cref{sec:TIS}\\
	& TIS estimation from BRDF ($\SI{1}{\degree}<\theta<\SI{75}{\degree})$ & 8 & $\pm$ 5 & \Cref{fig:BRDFdata}\\
	& TIS estimation from BRDF ($\SI{0.1}{\degree}<\theta<\SI{1}{\degree})$ & 4 & $\pm$ 5 & \Cref{fig:BRDFdata}\\	
	& TIS estimation from BRDF ($\SI{75}{\degree}<\theta<\SI{90}{\degree})$ & 0.16 & $\pm$ 0.4 & \Cref{fig:BRDFdata}\\ 
    & Total arm cavity loss from indirect measurements & 28 & $\pm$ 10 &  Main text \\
	\hline \hline
  \end{tabular}
  \caption{Summary of the experimental results. See main text for details. The TIS estimation has been split into three angular range : ($\SI{1}{\degree}<\theta<\SI{75}{\degree})$ allows direct comparison with TIS measurements, ($\SI{0.1}{\degree}<\theta<\SI{1}{\degree})$ uses the $\theta^{-2}$ dependence to estimate the angular range of missing data with the most uncertainty and ($\SI{75}{\degree}<\theta<\SI{90}{\degree})$ uses the flat 5 ppm loss in~\cite{Magana-Sandoval_2012_Large-Angle-Scatter}}
		
		\label{tab:paper_results}
\end{table*}

To compare the indirect measurement results to the direct ones, some highly non-trivial assumptions need to be made. First, we define scattering angles where less than 1\,ppm is reflected back from the mirror at the other end of the cavity as high-angle scattering. For the 40\,m arm cavities, the boundary is found to be $\sim\SI{0.065}{\degree}$. At scattering angles above this boundary, it is safe to use TIS to measure the loss. For scattering angles below that boundary, cavity simulation using phase maps are needed to calculate the round trip loss (see~\Cref{sec:phasemaps}). In our simulations, the maximum scattering angle is $\sim\SI{0.1}{\degree}$, above the required high-angle cutoff. Second, for high-angle scattering loss, we use measurement of other mirrors from the same vendor. The reason is that while specifics of the mirror coating greatly affect the low-angle scattering behavior, high-angle scattering is largely unchanged from that of the mirror substrate. Another consistency check that can be made by comparing the TIS estimated from the BRDF data shown in \Cref{fig:BRDFdata} and the TIS measured separately using an integrating sphere. The results, as can be seen in \Cref{tab:paper_results}, are consistent.

To obtain the roundtrip loss from the indirect measurements we sum together the following terms:
low-angle loss, IM average TIS measurement, EM average TIS measurement and 2$\times$ TIS estimation from the BRDF $(\SI{0.1}{\degree}<\theta<\SI{1}{\degree})$.
The result sums up to 28$ \pm $10\,ppm which comes close to the direct measurement of 35$ \pm $3\,ppm but has a large uncertainty, largely due to the uncertainty in the $\SI{0.1}{\degree}<\theta<\SI{1}{\degree}$ angular region. The careful measurement of that loss contribution should be a high priority for future studies.

To put our results in perspective, we plot the loss measurement per cavity length, the figure of merit for filter cavities \cite{Evans2013}, together with previous loss measurements of low-loss cavities \cite{Isogai_2013_LossStorageTime,DellaValle:14, Rempe:92, Ueda1996, kelles2007initial, Battesti2008, Sato:99} as a function of the equivalent confocal cavity length defined as
\begin{equation}
	\mathcal{L}_{\text{confocal}}=\frac{\pi w^2}{\lambda},
\end{equation}
where $w$ is the average beam waist on the optics and $\lambda$ is the wavelength of the input laser beam. The choice of equivalent confocal length instead of the actual cavity length is made to remove the dependence of the loss on the cavity geometry.

\section{Conclusions}
\label{s:Conclusion}
We have show here that our simulations accurately estimate the empirically measured optical losses
in the large beam regime (laser beam spot sizes $\gtrsim$1\,mm radius) which is relevant for long optical
cavities. In this regime, the losses are mainly due to figure errors of the mirror, rather than the microroughness. This work establishes that our simulations can be reliably utilized to predict the performance of large optical cavities, such as the ones used in the gravitional-wave detectors.

In the future, it will be important to make measurements in the small angle regime to accurately assess the losses which are due to coherent beaming from multiple point-defects.

\section*{Acknowledgments}
This work was supported by the National Science Foundation (NSF) operations grant PHY-1764464,
the Science and Technology Facilities Council under grants ST/V005693/1 and ST/V001019/1,
the Australian Research Council (ARC) through project numbers CE170100004 (Centre of Excellence for Gravitational Wave Discovery) and DE210100550, and the ANU Centre for Gravitational Astrophysics.
We would also like to acknowledge support from GariLynn Billingsley and Liyuan Zhang for supplying mirror surface maps and TIS measurements, and the many valuable conversations we have had with the Optics Working Group in the LIGO Scientific Collaboration. Advanced LIGO was built under Grant No. PHY-0823459. This material is based upon work supported by NSF’s LIGO Laboratory which is a major facility fully funded by the National Science Foundation.

\bibliographystyle{apsrev}
\bibliography{references}

\end{document}